\documentclass[fleqn,usenatbib]{mnras}
\usepackage{bm}
\usepackage[T1]{fontenc}
\DeclareRobustCommand{\VAN}[3]{#2}
\let\VANthebibliography\thebibliography
\def\thebibliography{\DeclareRobustCommand{\VAN}[3]{##3}\VANthebibliography}

\usepackage{graphicx}
\usepackage{amsmath}
\usepackage{amssymb}
\usepackage{xcolor}
\usepackage{ulem}

\usepackage{multirow}
\usepackage{booktabs}
\usepackage[flushleft]{threeparttable}
\usepackage{caption}

\def\lya{Lyman-$\alpha$}
\def\lyal{Lyman-$\alpha$ }

\def\hi{H$\,$\textsc{i} }

\def\h2{${\rm H}_2$}
\def\h2s{${\rm H}_2$}

\def\mathd{\mathrm{d}}

\def\nn{\nonumber}

\def\21cms{\textsc{21cmSense}}
\def\bifft{\textsc{BiFFT}}
\def\pyobs{\textsc{PyObs21}}

\def\mwa{\textsc{mwa}}

\def\hera{\textsc{hera}}

\def\skalow{\textsc{ska1-low}}

\newcommand{\uni}[2]{{\rm #1}^{#2}} 

\definecolor{notecolor}{rgb}{0.8,0,0}

\title[21-cm bispectrum at $z<6$]{The 21-cm bispectrum from neutral hydrogen islands at $z<6$}

\author[Raste et al.]{{Janakee Raste$^{1,2}$\thanks{E-mail: janakee@ncra.tifr.res.in},
  Girish Kulkarni$^2$,
  Catherine A.~Watkinson$^3$,
  Laura C.~Keating$^4$,}
  \newauthor{and Martin G.~Haehnelt$^{5,6}$}\\
  $^1$National Centre for Radio Astrophysics, Tata Institute of Fundamental Research, Pune 411007, India\\
  $^2$Tata Institue of Fundamental Research, Homi Bhabha Road, Mumbai 400005, India\\
  $^3$PolyChord Ltd, 27 Mortimer Street, London, W1T 3BL, UK \\
  $^4$Institute for Astronomy, University of Edinburgh, Blackford Hill, Edinburgh, EH9 3HJ, UK\\
  $^5$Institute of Astronomy, University of Cambridge, Madingley Road, Cambridge CB3 0HA, UK \\
  $^6$Kavli Institute of Cosmology, University of Cambridge, Madingley Road, Cambridge CB3 0HA, UK
}

\date{Accepted ---. Received ---; in original form ---}

\pubyear{2024}
\begin{document}
\label{firstpage}
\pagerange{\pageref{firstpage}--\pageref{lastpage}}
\maketitle

\begin{abstract}
  Spatial variations in the Lyman-$\alpha$ forest opacity at $z<6$ seem to require a late end to cosmic reionization.  In this picture, the universe contains neutral hydrogen `islands' of up to 100~cMpc$/h$ in extent down to redshifts as low as $z\sim 5.3$.  This delayed end to reionization also seems to be corroborated by various other observables.  An implication of this scenario is that the power spectrum of the cosmological 21-cm signal at $z<6$ is enhanced relative to conventional reionization models by orders of magnitude.  However, these neutral hydrogen islands are also predicted to be at the locations of the deepest voids in the cosmological large-scale structure.  As a result, the distribution of the 21-cm signal from them is highly non-Gaussian.  We derive the 21-cm bispectrum signal from these regions using high-dynamic-range radiative transfer simulations of reionization.  We find that relative to conventional models in which reionization is complete at $z>6$, our model has a significantly larger value of the 21-cm bispectrum.  The neutral islands also imprint a feature in the isosceles bispectrum at a characteristic scale of $\sim 1$~cMpc$^{-1}$.  We also study the 21-cm bispectrum for general triangle configuration by defining a triangle index.  It should be possible to detect the 21-cm bispectrum signal at $\nu\gtrsim 200$~MHz using \skalow\ for 1080 hours of observation, assuming optimistic foreground removal.
\end{abstract}


\section{Introduction}

\begin{figure*}
  \includegraphics[width=\linewidth]{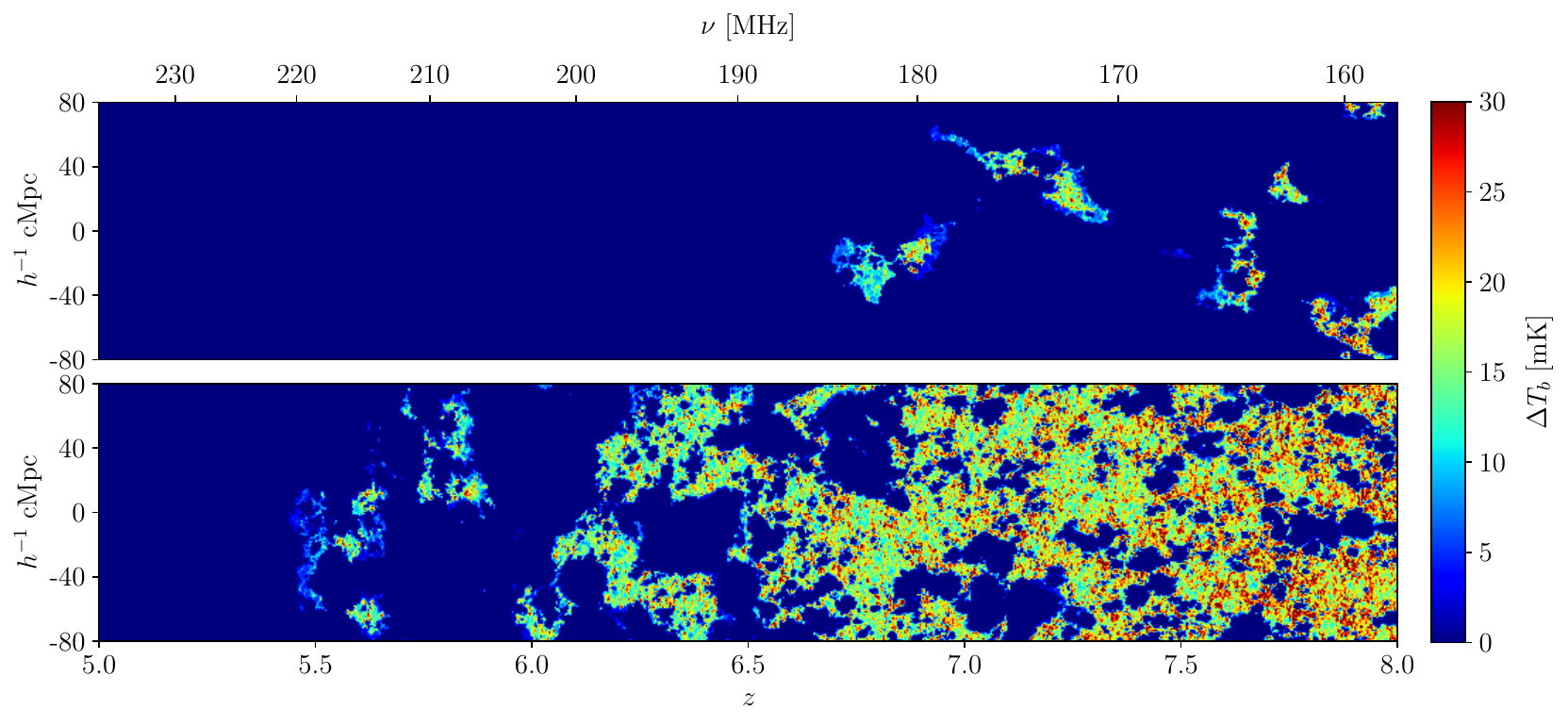}
  \caption{The bottom panel shows the 21-cm brightness temperature, $\Delta T_\mathrm{b}$, in a reionization model that is consistent with the Ly$\alpha$ forest at $z\sim 6$.  Large neutral hydrogen `islands' are seen at $z<6$.  In order to understand the 21-cm bispectrum signal from these neutral hydrogen islands, we contrast this reionization model with a more conventional one in which reionization finishes early, at $z>6$.  This model is shown in the top panel.  }
  \label{fig:lc}
\end{figure*}

Observations of the \lyal forest point to a late end of reionization \citep{2019MNRAS.485L..24K, 2020MNRAS.491.1736K, 2022MNRAS.514...55B}. In our previous work, we explored the implications of a reionization model that agrees with these observational constraints at redshifts 5--8 for the 21-cm power spectrum \citep{RK21}. We found that given the late end of reionization, the power spectrum of the 21-cm signal at redshifts $z=5$--$6$ is orders of magnitude higher than previous estimates.  This signal should be detectable by \hera\ and \skalow\ in $\sim 1000$ hours of observations, assuming optimistic foreground subtraction \citep{RK21}.  

However, models predict that the large islands of neutral hydrogen that persist till redshift $z\sim 5.5$ in our reionization models are in the deepest density voids in the universe.  As a result, the 21-cm signal from them should be significantly non-Gaussian, which should lead to a large bispectrum signal.  Furthermore, these neutral islands have highly irregular shapes that might hold clues about the galaxies that contributed to reionization.  While the power spectrum is sensitive to the size of the ionized regions, the bispectrum is sensitive to their shapes, which makes it a promising tool to study the topology of reionization \citep{2020MNRAS.492..653H}. Unlike the power spectrum, the value of the bispectrum signal for different triangle configurations can be negative as well as positive and this can encode information on various features of the reionization.  Simulations have shown that the modelling parameters of density, ionization, X-ray heating and \lyal coupling drive the non-Gaussianity at various scales. These processes determine the shape, magnitude, peak and sign of the 21-cm bispectrum as a function of redshift \citep{2016MNRAS.458.3003S,2018MNRAS.476.4007M, 2019MNRAS.482.2653W, 2020MNRAS.492..653H, 2021MNRAS.502.3800K, 2021ApJ...912..143M, 2021JCAP...12..024S, 2021arXiv210808201K}.  The 21-cm bispectrum is also a function of the redshift-space distortions and light-cone anisotropy \citep{2020MNRAS.493..594B, 2020MNRAS.499.5090M, 2021MNRAS.502.3800K, 2021JCAP...12..024S, 2021MNRAS.508.3848M}.  It has been consistently shown by multiple authors that observing the 21-cm bispectrum together with the power spectrum can reduce the inferred uncertainty on reionization parameters \citep{2016MNRAS.458.3003S, 2017MNRAS.468.1542S, 2022MNRAS.510.3838W, 2022JCAP...04..045T}. 

The 21-cm bispectrum signal can be observed by correlating the visibilities at three different baselines and frequencies \citep{2005MNRAS.358..968B, 2018PhRvL.120y1301T, 2020PhRvD.102b2001T, 2020PhRvD.102b2002T}.  The shape of the bispectrum triangle determines its detectability by interferometric experiments \citep{2021JCAP...12..024S, 2022JCAP...04..045T}. Particularly, the squeezed-limit isosceles bispectrum is expected to present the best observational prospects \citep{2019PASA...36...23T, 2022MNRAS.510.3838W, 2021MNRAS.508.3848M}.  Sensivity of bispectrum measurements, for radio-interferometric arrays, has also been explored  \citep{2015MNRAS.451..266Y, 2019MNRAS.487.4951S, 2021MNRAS.508.3848M}, for example for \skalow\ \citep{2022JCAP...04..045T, 2021MNRAS.508.3848M} and \mwa\ \citep{2019PASA...36...23T}.

In this paper we compute the bispectrum of the ionized hydrogen fraction, the gas density and the 21-cm brightness temperature for our late reionization model.  In Section~\ref{sec:simulation}, we briefly describe our simulation. We also discuss in this section the calculation of bispectra using the fast FFT code BiFFT presented by \cite{2017MNRAS.472.2436W} and various normalisations of the bispectrum. We present our results in Section~\ref{sec:results} for equilateral, isosceles and scalene triangle configurations. Finally, we discuss the prospects of detecting the 21-cm bispectrum with \skalow\ in Section~\ref{sec:sensitivity}, and conclude in Section~\ref{sec:conclusion}. 

We assume a flat $\Lambda$CDM universe with baryon and matter density parameters $\Omega_\mathrm{b}=0.0482$ and $\Omega_\mathrm{m}=0.308$, $\Omega_\Lambda=0.692$, Hubble constant $100~h\,\mathrm{km}\,\mathrm{s}^{-1}\,\mathrm{Mpc}^{-1}$ with $h=0.678$, spectral index of primordial curvature perturbations $n_\mathrm{s}=0.961$, clustering amplitude $\sigma_8=0.829$ at $z=0$, and helium mass fraction $Y_\mathrm{He}=0.24$ \citep{2014A&A...571A..16P}.  The units `ckpc' and `cMpc' refer to comoving kpc and comoving Mpc, respectively. 

\section{Methods} \label{sec:simulation}

\begin{figure*}
  \includegraphics[width=1\textwidth]{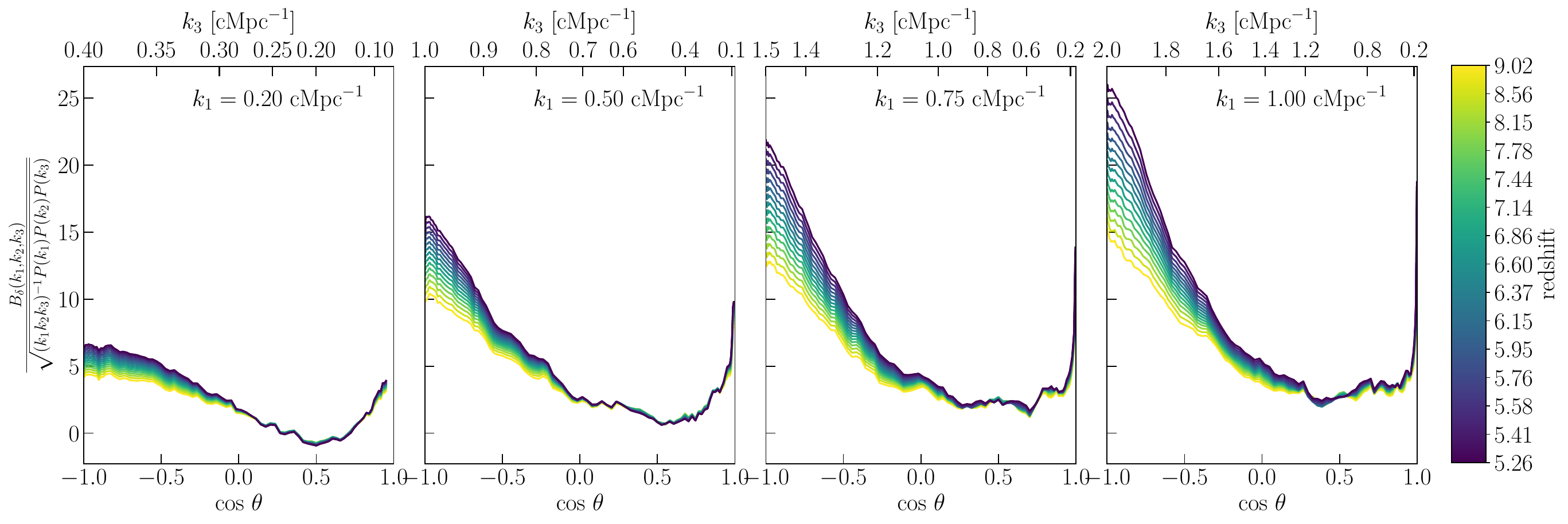}
  \caption{The normalised isosceles gas density bispectrum from redshift $z=9.02$ to $5.26$ for $k_1=0.2~{\rm cMpc}^{-1}$, $0.5~{\rm cMpc}^{-1}$, $0.75~{\rm cMpc}^{-1}$ and $1.0~{\rm cMpc}^{-1}$ (left to right). Amplitude of density bispectrum grows with time due to the increasing non-Gaussianity with the formation of structures. This amplitude also increases with $k_1$, as the small scales have more non-Gaussianity compared with larger scales. }
  \label{fig:isoc_D}
\end{figure*}

We have discussed our simulation in detail in \cite{2019MNRAS.485L..24K} and \cite{RK21}. Here we repeat only the essential details. 
To obtain the gas density and velocity fields, we have used the \textsc{p-gadget-3} code, a modified version of the \textsc{gadget-2} code \citep{2001NewA....6...79S, 2005MNRAS.364.1105S}. This simulation is similar to the simulations from the Sherwood Simulation Suite \citep{2017MNRAS.464..897B} with their 160--2048 initial conditions, containing $2048^3$ gas and dark matter particles and $160\,h^{-1}$~cMpc box length with periodic boundary conditions. For further processing, the gas density is gridded by projecting the smooth particle hydrodynamic (SPH) kernel in our simulation onto a Cartesian grid of size $2048^3$. This gives a grid resolution of $78.125\,h^{-1}\,\text{ckpc}$. The ionization and temperature fields are computed using the \textsc{aton} code \citep{2008MNRAS.387..295A, 2010ApJ...724..244A}, which solves the radiative transfer equation by using the M1 approximation for the first moment \citep{2008MNRAS.387..295A, 1984JQSRT..31..149L, 2008ASPC..385...91G}.

We calculate the differential brightness temperature ($\Delta T_\mathrm{b}$) box using the density, ionization, and peculiar velocity boxes and assuming $T_{\rm S} \gg T_{\rm CMB}$, 
\begin{align}
  &\Delta T_\mathrm{b}(\nu_o) \simeq 27 \; {\rm mK}\; x_{\rm HI} (1+\delta)  \left(1 + \frac{1}{H(z)} \frac{\mathd v_p}{\mathd s}\right)^{-1}\nn\\
  & \qquad\times\left(\frac{1+z}{10}\right)^{1/2}\left(\frac{Y_H}{0.76}\right)\left(\frac{0.14}{\Omega_m h^2}\right)^{1/2} \left(\frac{\Omega_b h^2}{0.022}\right).
\label{eq:21Tb_2}
\end{align}
We take the $z$-axis of the simulation box as the line-of-sight direction to calculate the peculiar velocity gradient $\mathd v_p/\mathd s$. The above expression assumes $\mathd v_p/\mathd s \ll H$, which is generally a valid assumption.  However, for the very few simulation cells in our computation that have that have $\mathd v_p/\mathd s\sim H$, we follow the standard practice and enforce a ceiling of $|\mathd v_p/\mathd s| < 0.5 H(z)$ \citep{2010MNRAS.406.2421S,21CMFAST,2012MNRAS.422..926M}.
%As our focus in this paper is on the end of reionization, at lower redshifts, we expect the effect of redshift space distortions to be small \citep{2013MNRAS.435..460J,2014MNRAS.443.2843M}. 

The reionization in our `late reionization' simulation model ends at redshift at $z\sim 5.3$, and the midpoint of reionization occurs at redshift $z \sim 7.1$. We also study an `early reionization' model, in which the evolution of the volume-averaged ionized hydrogen fraction is calibrated to match the evolution in the \cite{2012ApJ...746..125H} model of reionization. In this model, reionization is complete at $z\sim 6.7$. The two simulations are identical in all aspects apart from the source emissivity. Figure~\ref{fig:lc} compares the 21-cm brightness temperature in the two models. 

\subsection{Computing the Bispectrum} \label{sec:cal_bispctrum}
The bispectrum $B$ of a field $F(\bm{r})$ is defined as, 
\begin{multline}
	(2\pi)^3 B(\bm{k_1}, \bm{k_2}, \bm{k_3}) \delta_\mathrm{D}(\bm{k_1}+\bm{k_2}+\bm{k_3}) \\\equiv \langle \tilde{F}(\bm{k_1}) \tilde{F}(\bm{k_2}) \tilde{F}(\bm{k_3}) \rangle,
	\label{eq:BS_def}
\end{multline}
where $\tilde{F}(\bm{k})$ is the Fourier transform of $F(\bm{r})$. The Dirac delta $\delta_\mathrm{D}$ term requires that the wavevectors $\bm{k_1}$, $\bm{k_2}$ and $\bm{k_3}$ form a closed triangle in the Fourier space.

We follow  \citet{2015PhRvD..92h3532S}, \citet{2016MNRAS.460.3624S}, and \citet{2017MNRAS.472.2436W} to efficiently compute the bispectrum without using multiple nested loops through the Fourier box\footnote{In this work, we do not subtract the mean from the simulation box before calculating bispectra. All the information about the mean value is located only in the real part of the $\bm{k}=0$ mode of the Fourier box. This mode is not used while calculating power spectrum or bispectrum. We confirm that our results do not change by subtracting the mean from the simulation box.}.  This algorithm is described in detail by \citet{2017MNRAS.472.2436W} and is implemented by these authors in their publicly available code, \bifft\footnote{\url{https://bitbucket.org/caw11/bifft/}}.  We calculate bispectrum using a modified Python version of \bifft.  

\subsubsection{Triangle Configuration}

For any triangle formed by the wavevectors $\bm{k_1}$, $\bm{k_2}$ and $\bm{k_3}$,
\begin{align}
	k_3^2 = k_1^2 + k_2^2 - 2 k_1 k_2 \cos\theta,
\end{align}
where $\theta$ is the angle between $k_1$ and $k_2$ arms of the triangle.  For isosceles triangles, $k_1 = k_2$, so
\begin{align}
	\cos\theta = 1- \frac{k_3^2}{2k_1^2}.
\end{align}
Thus, in this case, for a fixed $k_1$, we can label a triangle equivalently using $\cos\theta$ or $k_3$. When $k_3 < k_1$, the angle $\theta<\pi/3$, and ${\rm cos}\;\theta > 0.5$.  Such triangles with $k_3\to 1$ and $\cos\theta \to 1$ are the so-called squeezed-limit triangles.  For equilateral triangles, $k_1 = k_2 = k_3$, so that $\theta = \pi/3$ and $\cos\theta = 1/2$.  Triangles with $k_3 > k_1$ have $\theta>\pi/3$, and $\cos\theta < 0.5$. The stretched-limit triangles have $\cos\theta\to -1$, with $k_3 \to 2k_1$.

\subsubsection{Bispectrum Normalization} \label{sec:norm}

Various normalizations of the bispectrum have been explored in the literature. In our work, we have either used the unnormalized bispectrum $B$ (Eq~\ref{eq:BS_def}), or a normalized bispectrum \citep{1968RvGSP...6..347H,1978PhFl...21.1452K, 1995ITSP...43.2130H, HINICH2005405, 2019MNRAS.482.2653W}, defined by
\begin{align}
	b = \frac{B(k_1, k_2, k_3)}{\sqrt{(k_1 k_2 k_3)^{-1}P(k_1)P(k_2)P(k_3)}}. \label{eq:norm}
\end{align}
The normalisation of the bispectrum isolates the non-Gaussianity of the field by removing the contributions of power spectrum. See \cite{2019MNRAS.482.2653W} for more discussion on various bispectrum normalisations. 
For the bispectrum of density and the neutral/ionized hydrogen fraction, the unnormalized bispectrum has units of ${\rm cMpc}^{6}$, whereas the unnormalised 21-cm brightness temperature bispectrum has units of ${\rm mK}^{3}\;{\rm cMpc}^{6}$. The normalised bispectrum from Eq~\ref{eq:norm} is dimensionless. 

\section{Results} \label{sec:results}

\begin{figure*}
  \includegraphics[width=1\textwidth]{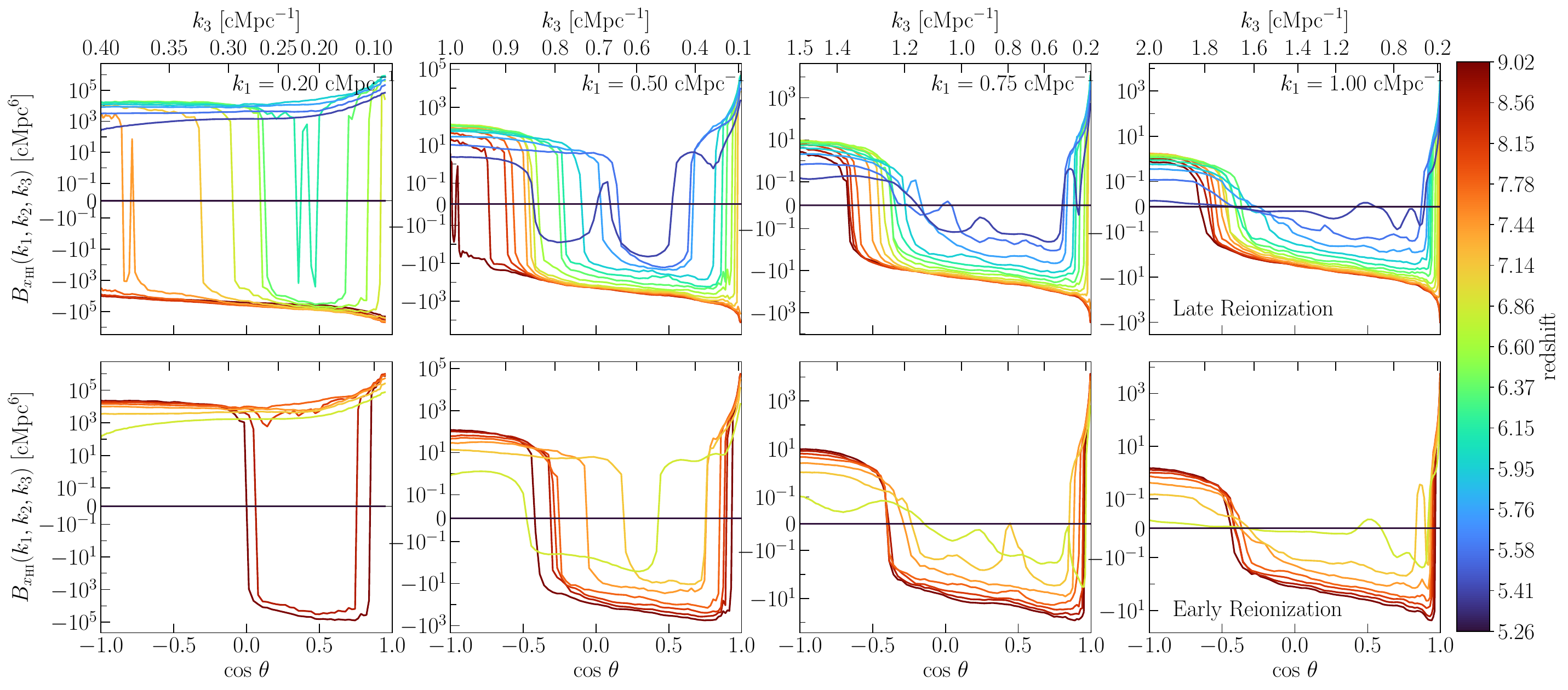}
  \caption{The un-normalized isosceles bispectrum of the neutral hydrogen fraction $x_\mathrm{HI}$ from redshift $z=9.02$ to $5.26$ at $k_1=0.2~{\rm cMpc}^{-1}$, $0.5~{\rm cMpc}^{-1}$, $0.75~{\rm cMpc}^{-1}$ and $1.0~{\rm cMpc}^{-1}$ (left to right) as function of $\cos\theta$ or $k_3$ in the late (top panel) and early (bottom panel) reionization models.}
  \label{fig:isoc_X_unnorm}
\end{figure*}

\begin{figure*}
  \includegraphics[width=1\textwidth]{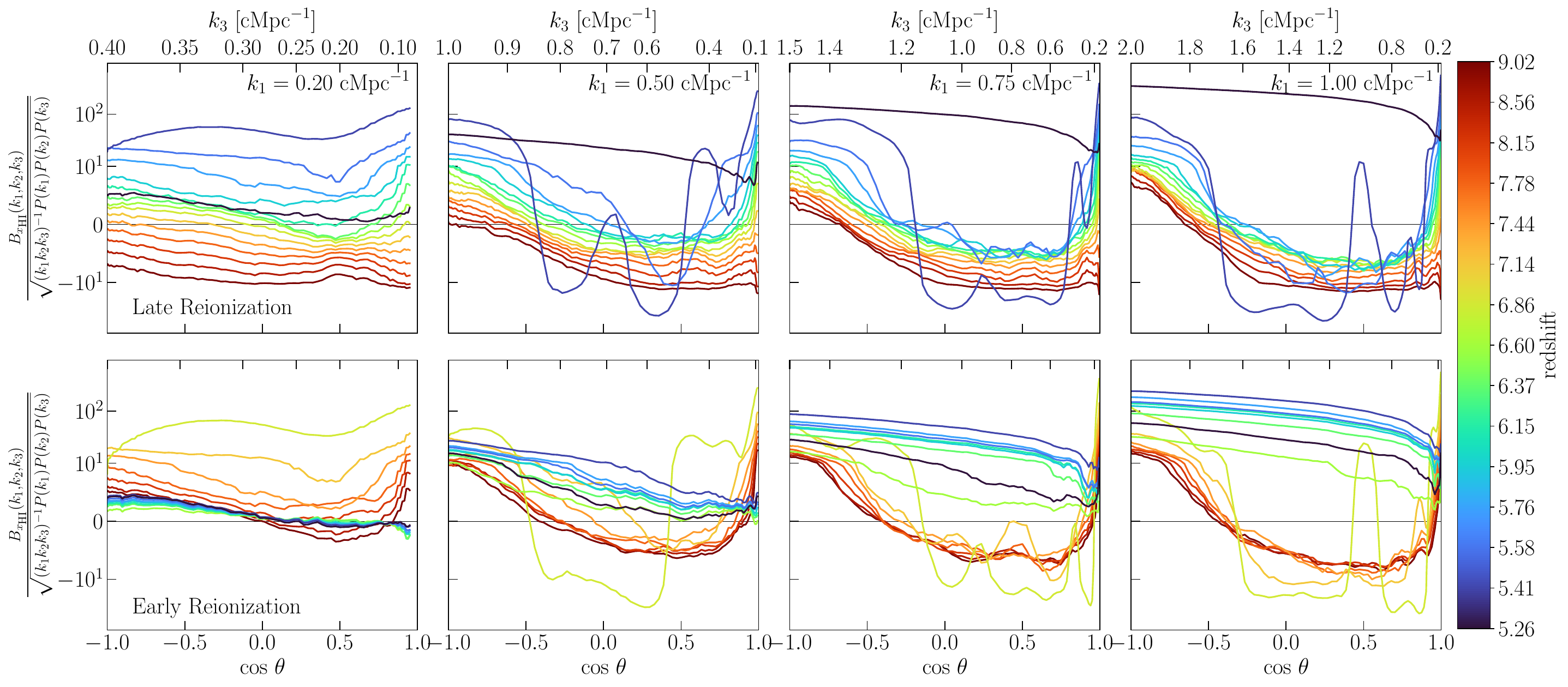}
  \caption{The normalized isosceles bispectrum of the neutral hydrogen fraction $x_\mathrm{HI}$ from redshift $z=9.02$ to $5.26$ at $k_1=0.2~{\rm cMpc}^{-1}$, $0.5~{\rm cMpc}^{-1}$, $0.75~{\rm cMpc}^{-1}$ and $1.0~{\rm cMpc}^{-1}$ (left to right) as function of $\cos\theta$ or $k_3$ in the late (top panel) and early (bottom panel) reionization models.}
  \label{fig:isoc_X}
\end{figure*}

\subsection{Density Bispectrum}

Figure~\ref{fig:isoc_D} shows the evolution of isosceles bispectra of the gas density from redshift $z=9.02$ to $5.26$. The isosceles triangle configuration is used for four representative values of $k_1=0.2, 0.5, 0.75$, and $1$~cMpc$^{-1}$.  For each value of $k_1$ ($=k_2$), the figure shows a range of values of $k_3$ available in the simulation box, between $0.08$ and $2$~cMpc$^{-1}$, depending on $k_1$.  We have normalised the bispectra using Equation~\ref{eq:norm} and the box has been reduced to resolution of $256^3$ for computational ease (see Appendix~\ref{sec:resolution} for a comparison with results from the higher resolution box).  The normalised density bispectrum is of the order of a few units, and grows with time due to an increase in the non-Gaussianity induced by structure formation.  The normalised bispectrum also increases in amplitude with increasing $k_1$, as the small scales have more non-Gaussianity compared with larger scales.

\begin{figure*}
 	\includegraphics[width=1\textwidth]{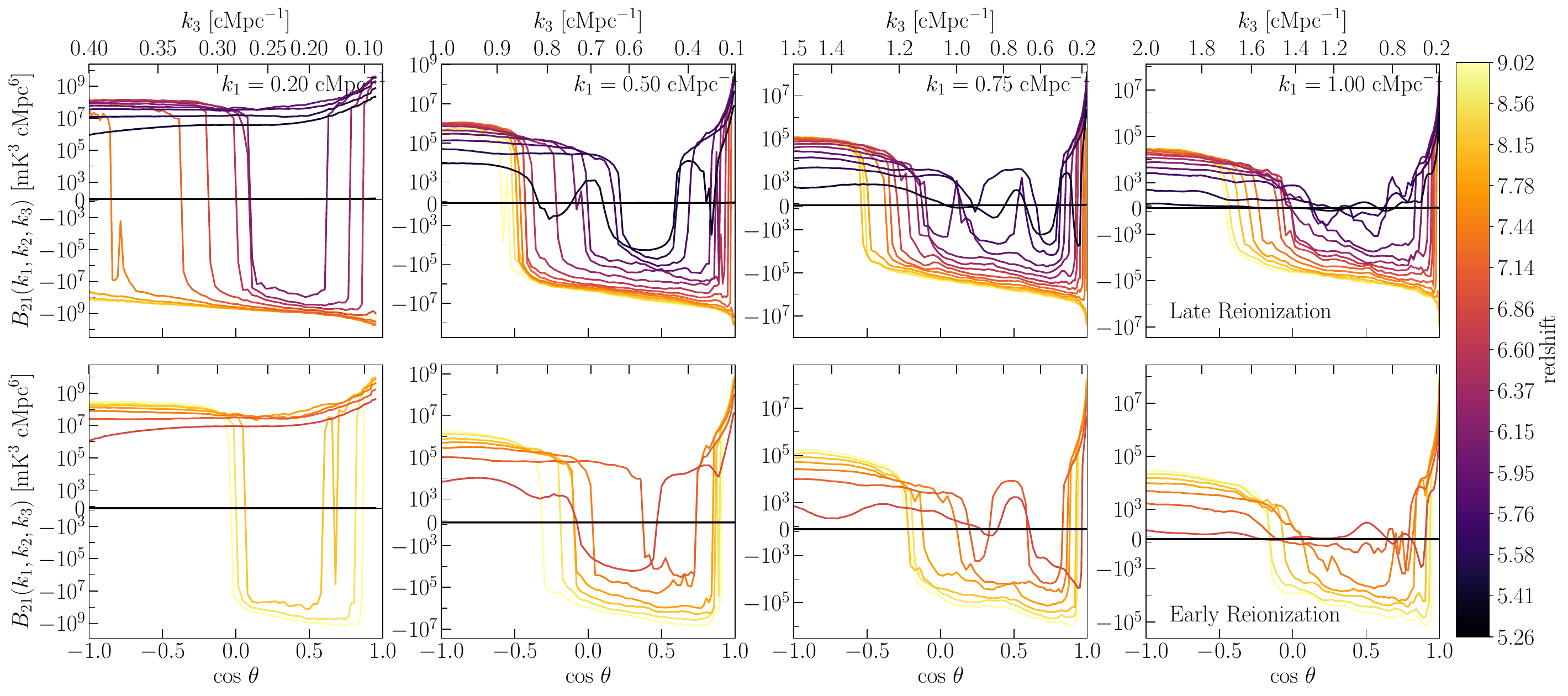}
  \caption{The un-normalized isosceles bispectrum of the 21-cm brightness temperature ($\Delta T_\mathrm{b}$) from redshift $z=9.02$ to $5.26$ at $k_1=0.2~{\rm cMpc}^{-1}$, $0.5~{\rm cMpc}^{-1}$, $0.75~{\rm cMpc}^{-1}$ and $1.0~{\rm cMpc}^{-1}$ (left to right) as function of $\cos\theta$ or $k_3$ for late (top panel) and early (bottom panel) reionization models.}
  \label{fig:isoc_B_unnorm}
\end{figure*}

\begin{figure*}
 	\includegraphics[width=1\textwidth]{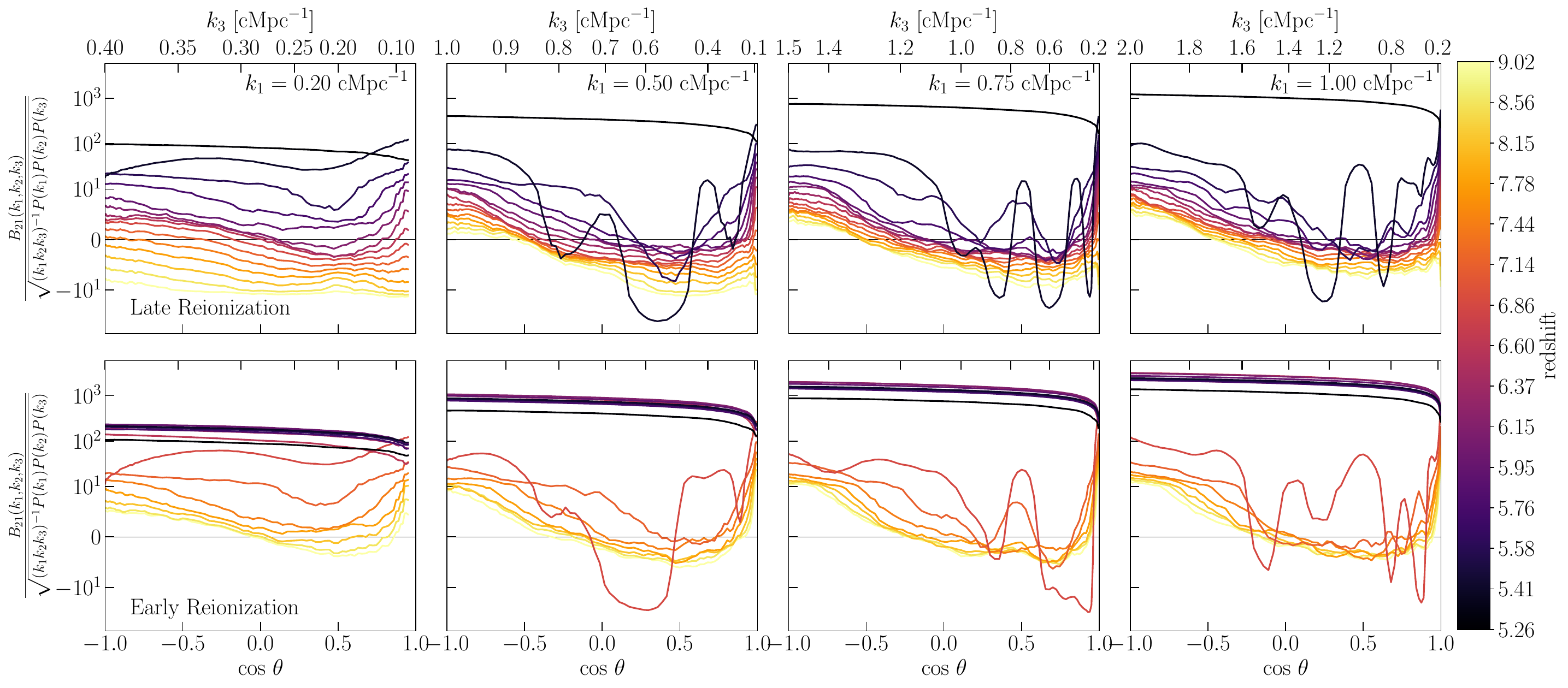}
  \caption{The normalized isosceles bispectrum of the 21-cm brightness temperature ($\Delta T_\mathrm{b}$) from redshift $z=9.02$ to $5.26$ at $k_1=0.2~{\rm cMpc}^{-1}$, $0.5~{\rm cMpc}^{-1}$, $0.75~{\rm cMpc}^{-1}$ and $1.0~{\rm cMpc}^{-1}$ (left to right) as function of $\cos\theta$ or $k_3$ for late (top panel) and early (bottom panel) reionization models.}
  \label{fig:isoc_B}
\end{figure*}

\subsection{Neutral Hydrogen Fraction Bispectrum}

Figures~\ref{fig:isoc_X_unnorm} and \ref{fig:isoc_X} show, respectively, the unnormalised and normalised bispectrum of the neutral hydrogen fraction at various redshifts between $z=9.02$ and $5.26$, at $k_1=0.2, 0.5, 0.75$ and $1$ cMpc$^{-1}$.  Each of the two figures show the bispectrum for both of our reionization models in separate panels. Apart from the difference in the redshift of reionization, the bispectrum in the two reionization models are qualitatively similar, only shifted in redshift (time).

\begin{figure*}
\centering
\includegraphics[width=0.49\linewidth]{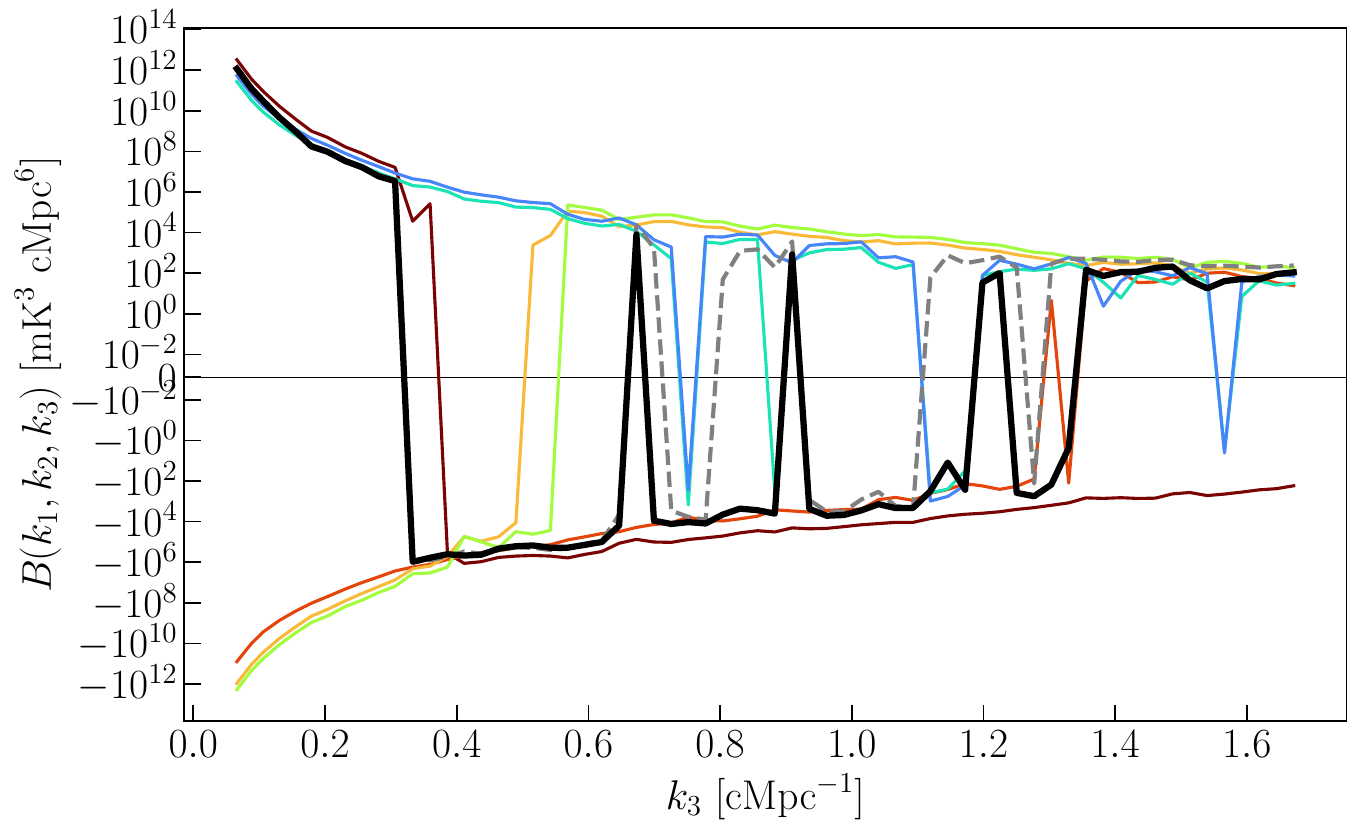}
\includegraphics[width=0.49\linewidth]{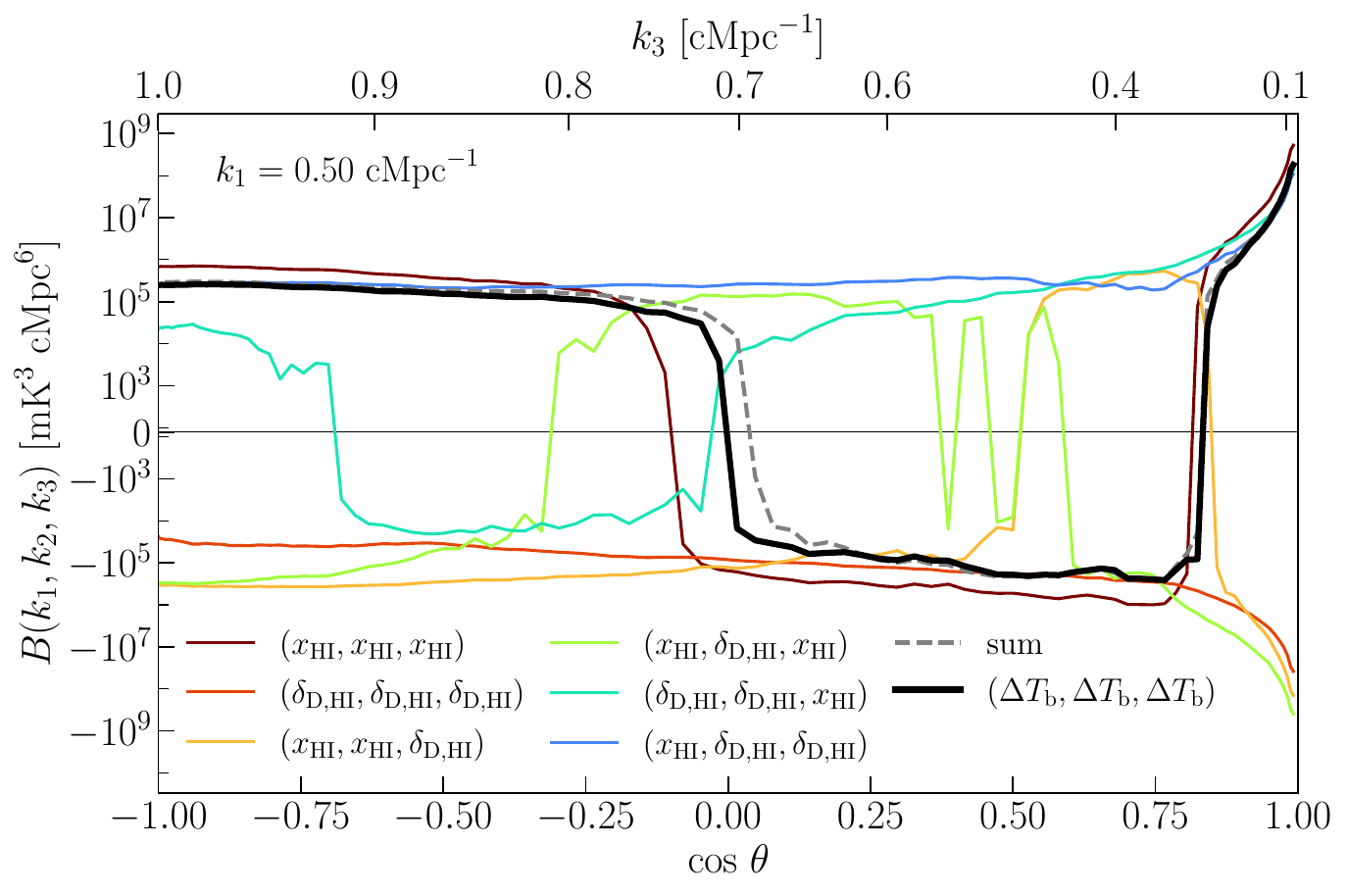}
\caption{Unnormalised cross bispectra components, multiplied by their respective weights, from Eq~\ref{eq:cross_terms_neutral} for equilateral (left) and isosceles (right, at $k_1 = 0.5~{\rm cMpc}^{-1}$) bispectrum configuration at $z=5.95$ for the late reionization model. We compare the weighted sum (gray dashed curves) of these components with the $\Delta T_\mathrm{b}$ bispectra (thick black curves).
}
\label{fig:cross_neut}
\end{figure*}

Recall that $\cos\theta=0.5$ denotes the equilateral triangle configuration. Values of $\cos\theta$ from $-1$ to $0.5$ correspond to the stretched limit, and those from $0.5$ to $1$ correspond to the squeezed limit.  A large positive value of the equilateral bispectrum indicates an overabundance of roughly spherical structures of higher-than-average values of $x_\mathrm{HI}$ (neutral regions) embedded in the background of lower-than-average values of $x_\mathrm{HI}$ (ionized regions) \citep{2011JCAP...10..026L,2020MNRAS.492..653H}.  A large negative value indicates an over-abundance of below-average structures embedded in the above-average background. This allows us to interpret the evolution that we see in Figure~\ref{fig:isoc_X_unnorm}.  

At high redshifts, the $x_\mathrm{HI}$ distribution has `holes' of below-average values (the ionized regions), yielding a negative values for almost all $k$ modes and triangle configurations. However, the bispectra at very small scale (large $k$) stretched limit triangles are positive. These triangles correspond to over-abundance of small-scale above-average filamentary structure. This is perhaps the small scale neutral `valleys' between spherical ionized bubbles.

With the progress of reionization, the ionized regions become larger and start merging. As reionization crosses the half-way point, the distribution of $x_\mathrm{HI}$ now has `islands' of above-average values, yielding a positive value of the bispectrum. At the mid-point of reionization, even if roughly half of the volume is occupied by ionized IGM and half by neutral, the shapes of ionized regions are roughly spherical compared to neutral regions, which have more irregular shapes. Therefore, we see that the stretched and squeezed limit bispectrum start becoming positive at lower redshifts, however small scale (large $k$) equilateral bispectra still remain negative. They are signature of small scale spherical ionized regions embedded in neutral IGM. This signature persists even during the later half of reionization. At large scales (small $k$), on the other hand, the bispectrum is positive for all configurations during the later part of reionization. This suggests that, on large scales, now there are above-average neutral regions embedded in below-average ionized IGM.

In the stretched limit, $\cos\theta\rightarrow -1$, the bispectrum measures the probability of the neutral hydrogen pancakes that demarcate ionized bubbles.  Such pancake-like boundary surfaces are the over-abundant structure at all redshifts.  Consequently, the bispectrum for $\cos\theta\rightarrow -1$ is positive at all redshifts after the start of reionization.

In the squeezed limit, $\cos\theta\rightarrow 1$, a positive value of the bispectrum indicates an overabundance of small-scale positive perturbations in the $x_\mathrm{HI}$ distribution and a large-scale modulation of these perturbations. As soon as the ionized regions grow to a reasonable size, the distribution of $x_\mathrm{HI}$ inside the large ionized regions is trivially uniform, but that in the leftover neutral regions has small-scale perturbations due to smaller ionized bubbles.  The bispectrum turns positive when this happens.

At the end of reionization, at redshifts of $z\sim 5.4$, the bispectrum shows complex features that are sensitive to the morphology of the residual $x_\mathrm{HI}$ islands. In Figure~\ref{fig:isoc_X}, the normalisation accentuates the features in the bispectrum that we see in Figure~\ref{fig:isoc_X_unnorm}. This is most pronounced for the oscillating features at redshift $z \sim 5.4$. These oscillations are a qualitatively distinct signature that seems to mark the end phases of reionization in both early and late reionization models, and so could be a useful smoking gun for identifying the redshift of reionization from observations.

\begin{figure*}
  \includegraphics[width=1\textwidth]{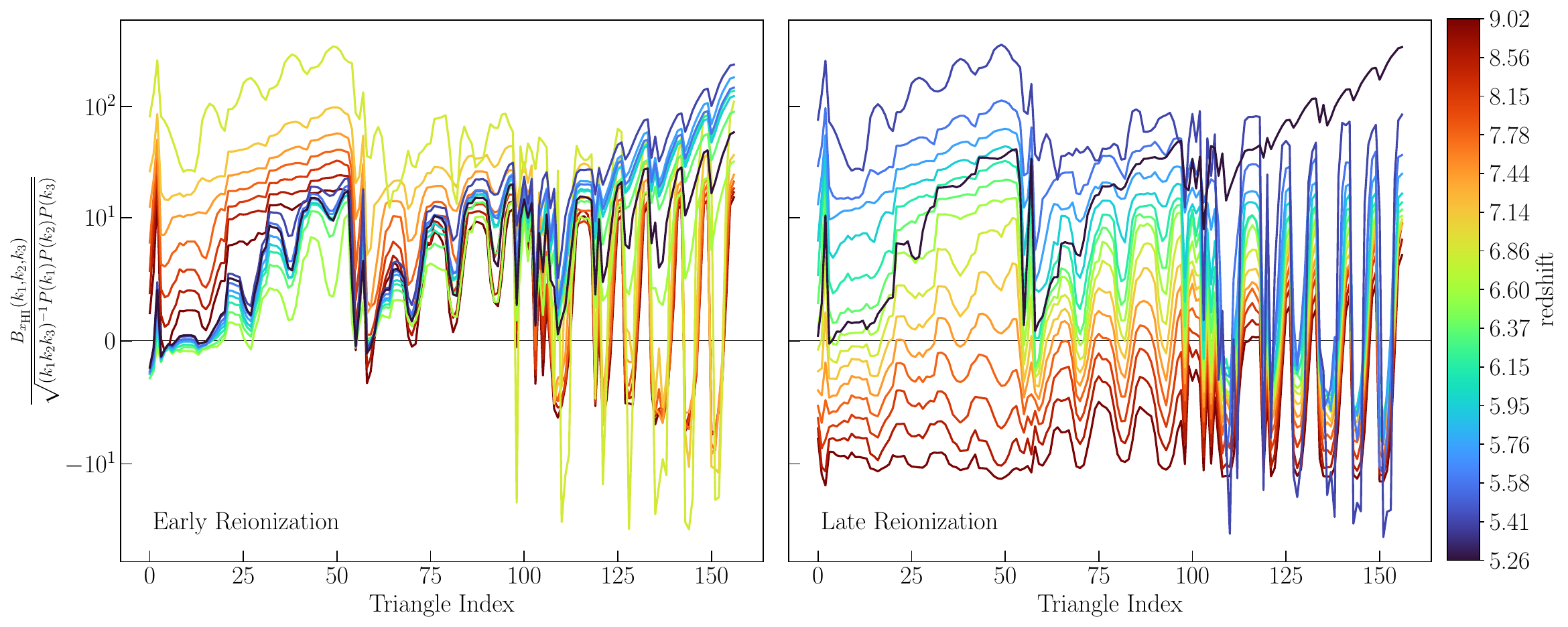}
  \includegraphics[width=1\textwidth]{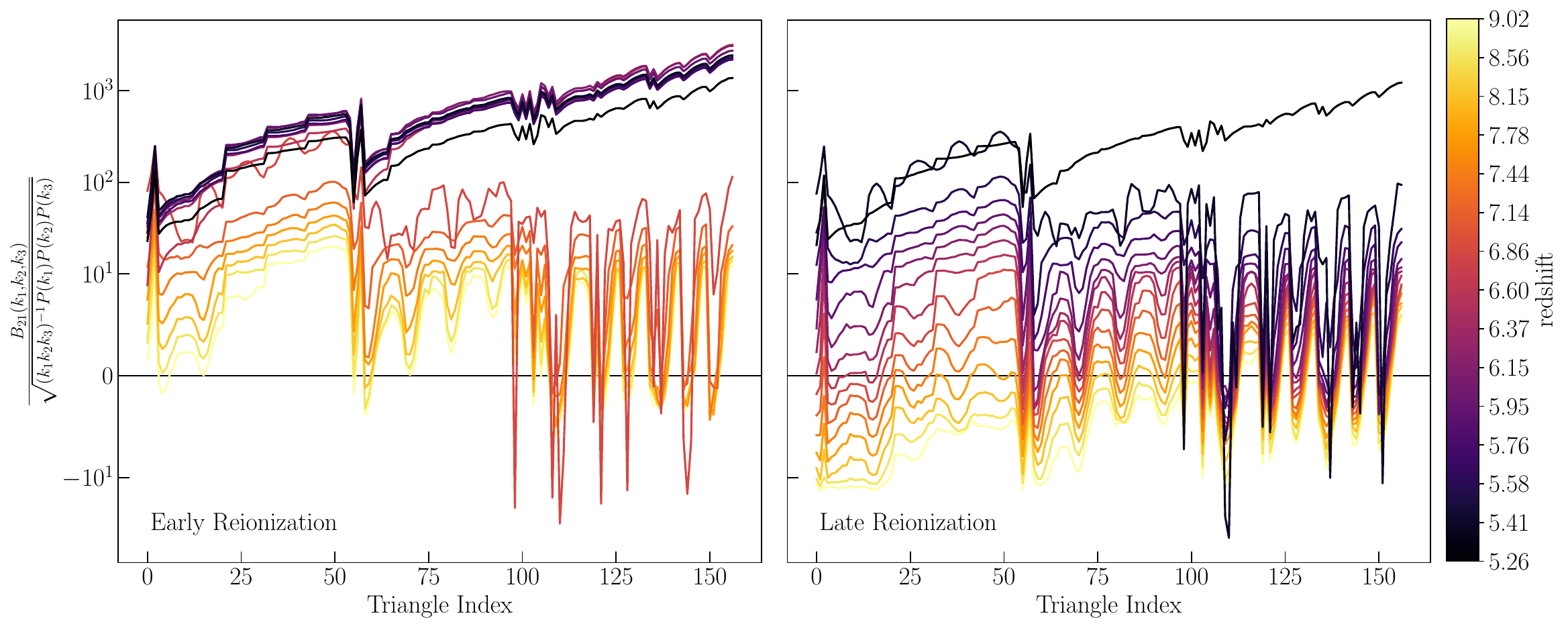}
  \caption{Normalized bispectrum for neutral fraction (top panel) and brightness temperature (bottom panel) for the late (right panel) and early (left panel) reionization models from redshift $z=9.02$ to $5.26$ for all triangle configurations (see Table~\ref{tab:TriangleIndices}).}
  \label{fig:gen_1}
\end{figure*}

\subsection{21-cm Bispectrum}

Figures~\ref{fig:isoc_B_unnorm} and \ref{fig:isoc_B} respectively show the unnormalised and normalised isosceles bispectrum of the 21-cm brightness temperature for the same redshift range and the $k$ values as in Figures~\ref{fig:isoc_X_unnorm} and \ref{fig:isoc_X}. For approximately equilateral triangles, the bispectrum is negative in the early stages of reionization, changes sign when reionization is half complete, and then is positive at lower redshift, before settling to zero in the post-reionization era.  The bispectrum is positive at almost all redshifts in the stretched limit.  Similarly, the bispectrum in the squeezed limit also stays positive at all but the very early stages of reionization.  At the end of reionization, for $z\sim 5.4$, the bispectrum shows a set of complex features that map to the 21-cm brightness structure of the residual neutral hydrogen islands in the voids. 

It is noteworthy that, similar to the 21-cm power spectrum explored in our previous work \citep{RK21}, the predicted bispectrum is very large at $z<6$ in the late reionization model, while the bispectrum at these redshifts is zero in the fiducial early reionization model.  This is due to the persistence of neutral hydrogen islands at these redshift in the late reionization model.  This signal in the 21-cm bispectrum is directly induced by the opaque regions seen in the Ly$\alpha$ forest at these redshifts.  

In the absence of spin temperature fluctuations, most of the brightness temperature bispectrum is induced by the fluctuations in $x_\mathrm{HI}$ in the range of redshifts and wavenumbers considered here. Therefore, we see that brightness temperature bispectrum follows a very similar trend as the bispectrum of the neutral hydrogen fraction.  
To study this effect more quantitatively, we break down our $\Delta T_\mathrm{b}$ bispectra in various auto- and cross-bispectra components of density and neutral fraction. Ignoring the redshift space distortion, the 21-cm signal at any redshift can be written as, 
\begin{align}
    \Delta T_\mathrm{b}(z) = (1+\delta_\mathrm{D}) x_{\rm HI} T_0(z),
    \label{eq:Tb_components}
\end{align}
where, $T_0(z)$ is the base 21-cm signal at redshift $z$ and there are spatial fluctuations due to density ($\delta_D$) and neutral \hi fraction ($x_{\rm HI}$). The average 21-cm signal is,
\begin{align}
    \overline{\Delta T_\mathrm{b}} = T_0 \langle (1+\delta_\mathrm{D}) x_{\rm HI} \rangle.
\end{align}
And its fluctuation is,
\begin{align}
    \delta_\mathrm{T_\mathrm{b}} \overline{\Delta T_\mathrm{b}} &= \Delta T_\mathrm{b} - \overline{\Delta T_\mathrm{b}} \nn \\
    &= T_0 [x_{\rm HI}+\delta_\mathrm{D}x_{\rm HI}]  - T_0 \left[ \langle x_{\rm HI} \rangle + \langle \delta_\mathrm{D} x_{\rm HI}  \rangle \right] \nn \\
    &= T_0 [(x_{\rm HI} - \langle x_{\rm HI} \rangle) + (\delta_\mathrm{D} x_{\rm HI}  - \langle \delta_\mathrm{D} x_{\rm HI}  \rangle) ].
\end{align}
Defining $\delta_{\rm HI} = (x_{\rm HI} - \langle x_{\rm HI} \rangle) /\langle x_{\rm HI} \rangle $ and $\delta_{\rm D,HI} = (\delta_\mathrm{D} x_{\rm HI} - \langle \delta_\mathrm{D} x_{\rm HI} \rangle) /\langle \delta_\mathrm{D} x_{\rm HI} \rangle $, we have, 
\begin{align}
    (\overline{\Delta T_\mathrm{b}})^3 \langle \delta_\mathrm{T_\mathrm{b}} \delta_\mathrm{T_\mathrm{b}} \delta_\mathrm{T_\mathrm{b}} \rangle &= T_0^3 \langle [\overline{ x_{\rm HI}} \delta_{\rm HI} + \overline{ \delta_\mathrm{D} x_{\rm HI} } \delta_{\rm D,HI} ]^3 \rangle \nn \\
    &= T_0^3 \langle (\overline{ x_{\rm HI}} \delta_{\rm HI})^3 + (\overline{ \delta_\mathrm{D} x_{\rm HI} } \delta_{\rm D,HI})^3 \nn \\
        & \quad + 3 \overline{ x_{\rm HI}}^2 \overline{ \delta_\mathrm{D} x_{\rm HI}} \delta_{\rm HI}^2 \delta_{\rm D,HI}\nn \\
        & \quad + 3 \overline{ x_{\rm HI}} \overline{ \delta_\mathrm{D} x_{\rm HI}}^2 \delta_{\rm HI} \delta_{\rm D,HI}^2 \rangle.
        \label{eq:cross_terms_neutral}
\end{align}
Notice that the density bispectrum does not contribute directly to the $\Delta T_\mathrm{b}$ bispectra. We should also emphasise that while calculating cross bispectra, the order of the fields is important in a non-equilateral configuration. For example, $\langle \delta_{\rm D} \delta_{\rm HI} \delta_{\rm HI}\rangle$ might not be the same as $\langle \delta_{\rm HI} \delta_{\rm D} \delta_{\rm HI} \rangle$ if $\bm{k}_1 \neq \bm{k}_2$.

\begin{figure*}
\centering
\includegraphics[width=0.95\linewidth]{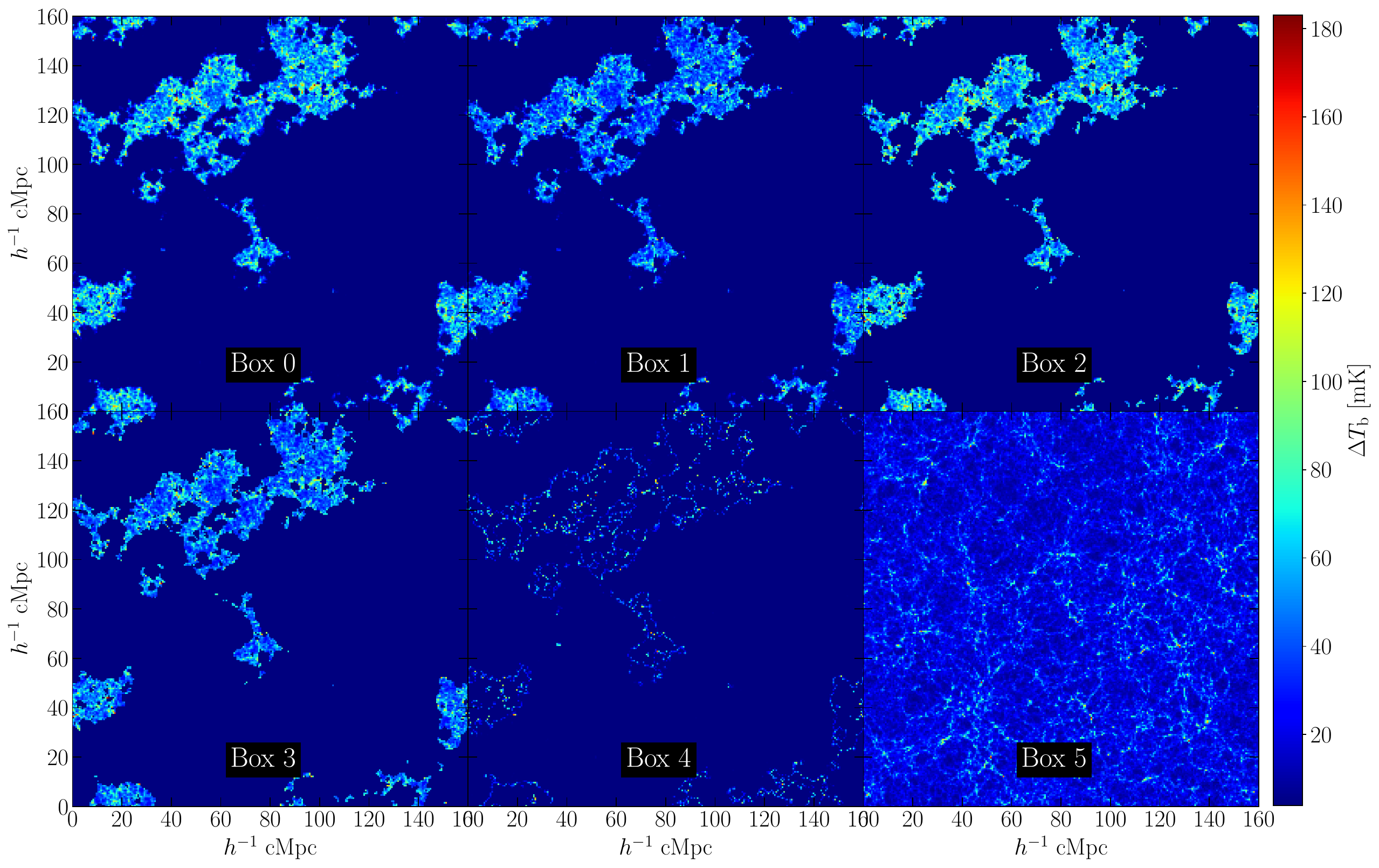}
\caption{$\Delta T_\mathrm{b}$ boxes 0 to 5 for redshift $z = 5.96$ for the late reionization model. These boxes were computed after modifying the ionization field by hand. Box 0 is the original simulations box, whereas in Box 1, $x_{\rm HII} < 0.5$ regions are set to $x_{\rm HII} = 0$, and in Box 2, $x_{\rm HII} \geq 0.5$ regions are set to $x_{\rm HII} = 1$. Box 3 has both these approximations, essentially converting the box to a binary field of 0 and 1. Finally, in Box 4 $x_{\rm HII} \leq 0.5$ regions are set to $x_{\rm HII} = 1$
and in Box 5 $x_{\rm HII} \geq 0.5$ regions are set to $x_{\rm HII} = 0$, which respectively removes neutral and ionized regions from the simulation box.}
\label{fig:partial_ion_box_2}
\end{figure*}

\begin{figure*}
  \centering
  \includegraphics[width=1\linewidth]{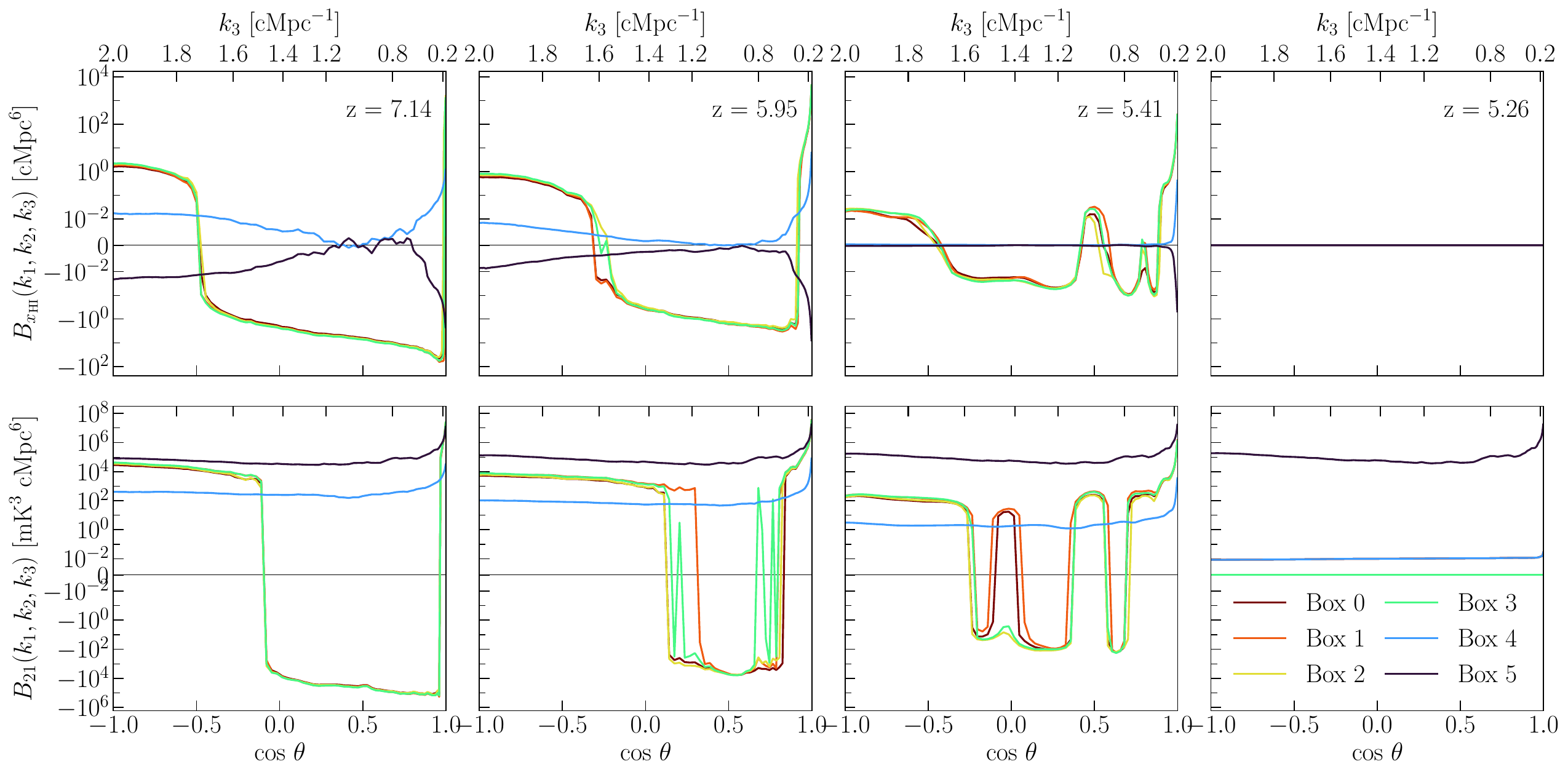}
  \caption{Unnormalised isosceles bispectra of $x_{\rm HI}$ (top) and $\Delta T_\mathrm{b}$ (bottom) for the late reionization model at $k_1 = 1~\uni{cMpc}{-1}$ for boxes 0 to 5 and redshifts 7.14, 5.95, 5.41 and 5.26 from left to right. Boxes 0 to 3 have very slight differences, which suggests that perhaps, the bispectra are more dependent on the neutral vs ionized state of the IGM, and the partially ionized regions do not affect the bispectra significantly. Box~4 has large, flat, positive $\Delta T_{\rm b}$ bispectra at all redshifts, which matches with the post-reionization bispectra in the last panel. In this box, we have removed the effect of neutral regions. Therefore, the post-EoR bispectra is an effect of residual neutral fraction, as expected. $\Delta T_{\rm b}$ bispectrum of Box~5 is in essence the gas density bispectrum.}  
  \label{fig:neutral_islands1_costheta}
\end{figure*}

In Figure~\ref{fig:cross_neut} we show each of the cross bispectra (unnormalised) component, multiplied by its respective weight, from Eq~\ref{eq:cross_terms_neutral} for the equilateral bispectrum configuration (left) and the isosceles configuration at $k_1 = 0.5~\uni{cMpc}{-1}$, for redshift $z=5.95$. The gray dashed curve shows the weighted sum of these individual components. Comparing it with the $\Delta T_\mathrm{b}$ bispectra, we see that this predicted bispectrum is slightly different from the computed bispectra, since we have ignored here the effect of velocity fluctuations in Eq~\ref{eq:Tb_components}. But this effect is small. We also note that the bispectra of various cross components fluctuate around zero a lot more than the auto bispectra. These fluctuations occur where the non-Gaussianity is close to 0. However, the sum of these cross components do not show these fluctuations, as their effects average out. Finally, note that the $\Delta T_\mathrm{b}$ bispectra have shapes very similar to the neutral fraction bispectra. Hence, in absence of spin temperature fluctuations, the neutral fraction fluctuations dominate the 21-cm fluctuations. 

Other than a few $k$-modes, which show the positive fluctuations, the equilateral bispectrum of $\Delta T_{\rm b}$ is negative at intermediate $k$-modes and postive at very large and very small $k$-modes. Also notice that for the equilateral configurations $\langle\delta_{\rm HI} \delta_{\rm HI} \delta_{\rm D,HI} \rangle$ and $\langle \delta_{\rm HI} \delta_{\rm D, HI} \delta_{\rm HI} \rangle$, as well as $\langle \delta_{\rm D, HI} \delta_{\rm D, HI} \delta_{\rm HI}\rangle$ and $\langle \delta_{\rm HI} \delta_{\rm D, HI} \delta_{\rm D, HI} \rangle$ have very similar shapes, other than few fluctuations. However, their shapes are very different for the isosceles configurations.

\begin{figure*}
    \centering
    \includegraphics[width=1\linewidth]{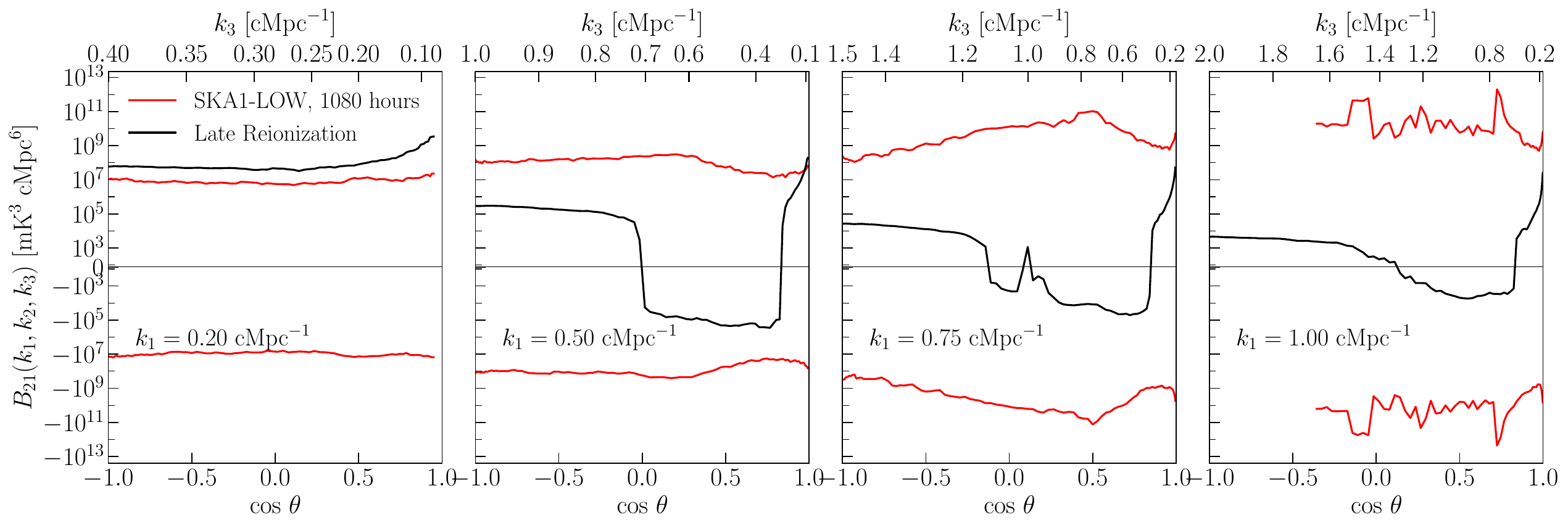}
    \caption{\skalow\ sensitivity (red curves) for 1080 hours of tracking mode observation with optimistic foreground removal at redshift $z= 5.96$ ($\nu = 204~{\rm MHz}$) and $k_1 = $ 0.2, 0.5, 0.75 and 1.0$~\uni{cMpc}{-1}$ from left to right. We compare it with our late reionization model (black curves). The positive and negative sensitivity curves are the 1-$\sigma$ upper and lower bounds for the real part of noise bispectra.}
    \label{fig:sense1}
\end{figure*}

\subsection{Bispectrum for generic triangle configurations}

Moving from the isosceles triangle to more general triangle configurations, in Figure~\ref{fig:gen_1}, we show the evolution of the normalized bispectra for available triangle configurations for the $256^3$ resolution cube of neutral fraction (top panels) and brightness temperature (bottom panels), in the late (right) and early (left) reionization models.

The mapping between the triangle index and $k$ values is as follows: Of the ($k_1$, $k_2$, $k_3$) triplet, two values are taken from array (0.1, 0.2, 0.5, 0.75, 1.0~$\uni{cMpc}{-1}$), with $k_1 \geq k_2$. We choose the third $k$ value with $\theta/\pi$ from the array [0.01, 0.05, 0.1, 0.2, 0.33, 0.4, 0.5, 0.6, 0.7, 0.85, 0.95].
Then, ($k_1$, $k_2$, $k_3$) triplets are sorted in increasing order ($k_a$, $k_b$, $k_c$) with $k_a \geq k_b \geq k_c$. The triangle index of a given triplet is its rank in the sorted sequence, with $k_a$ moving fastest. We tabulate the mapping between the triangle index and the ($k_1$, $k_2$, $k_3$) triplets in Appendix~\ref{sec:triangleindex}. 
Indices $> 100$ roughly correspond to $k>0.5~\uni{cMpc}{-1}$.
The fluctuations at higher triangle indices are due to the triangle configuration fluctuating between stretched limit triangles, which correspond to the large positive bispectra values and other configurations, including equilateral, which correspond to negative bispectra values. 

We can see that for early stages of reionization, the bispectra of neutral fraction for various triangle configurations is negative. With the evolution of reionization, they become positive. Towards the end of reionization, the amplitude of the fluctuations increase. The post-reionization bispectra have small amplitude for small triangles (small $k$-modes, large scale), but they become larger for larger triangle configurations (large $k$-modes, small scale). The brightness temperature bispectra have very similar behaviour, however, their amplitude is slightly more positive, reflecting the effect of density bispectra. The post-reionization normalised $\Delta T_\mathrm{b}$ bispectra are usually larger than the reionization bispectra at all $k$-mode triangles and do not show strong fluctuations with triangle index.

\subsection{Untangling the bispectrum}

To understand which features of our simulation box correspond to which details in the 21-cm bispectrum, we take our original ionization fraction box (Box 0) and construct several different modified boxes using following prescription:
\begin{itemize}
\item Box 0: Original
\item Box 1: $x_{\rm HII} < 0.5$ is set to $x_{\rm HII} = 0$
\item Box 2: $x_{\rm HII} \geq 0.5$ is set to $x_{\rm HII} = 1$
\item Box 3: Both of the above (essentially converting the box to a binary field of 0 and 1)
\item Box 4: $x_{\rm HII} \leq 0.5$ is set to $x_{\rm HII} = 1$
\item Box 5: $x_{\rm HII} \geq 0.5$ is set to $x_{\rm HII} = 0$
\end{itemize}

In Figure~\ref{fig:partial_ion_box_2}, we show slices of brightness temperature boxes created using these modified ionization boxes at redshift $z = 5.96$. We show unnormalised isosceles bispectra for these boxes at $k_1 = 1~\uni{cMpc}{-1}$, for various redshifts in Figure~\ref{fig:neutral_islands1_costheta}. Boxes 0 to 3 have very slight differences. Specifically, we can see that box 0 and box 3 have very similar shape and magnitude for almost all redshifts and almost all $k$-modes. This suggests that perhaps, the bispectra are more dependent on the neutral vs ionized state of the IGM, and the partially ionized regions do not affect the bispectra significantly. However, when we remove these ionized shapes and regions from the simulation box in Box~4, the shape of the bispectrum changes completely. Boxes~4 and~5 are mostly ionized and neutral boxes, respectively, with only some partially ionized regions left. Boxes 4 and 5 provide useful extreme cases with which to compare and contrast boxes 0--3.  Box~5 represents a nearly completely neutral volume; only small partially ionized regions exist in an otherwise neutral IGM.  Similarly, box~4 represents a nearly completely ionized volume; only small partially neutral regions are present in an otherwise ionized box.  A comparison of such configurations with boxes 0--3 isolates contribution to the bispectrum of the most visible component of boxes 0--3, namely the large residual neutral regions.  Indeed, the bispectra of boxes 4 and 5 are qualitatively different from the bispectra in boxes 0--3.  The ionization fraction bispectra for Box~4 and Box~5 have similar shape with the sign inverted, as we see in top panels of Figure~\ref{fig:neutral_islands1_costheta}. The bispectra also have a smaller amplitude.  The 21-cm bispectra are positive in boxes 4 and 5.  In box 5, the bispectrum is set by density fluctuations.  These do not evolve significantly from redshift 7 to 5, leading to the almost negligible change with redshift in the bispectrum in the bottom panels of Figure~\ref{fig:neutral_islands1_costheta}.  Box~4 on the other hand shows a bispectrum that is set by the small number of partially ionized cells.  The resultant bispectra matches the post-reionization values that we see in Figure~\ref{fig:isoc_B_unnorm}.  The redshift evolution seem in the 21-cm bispectrum in box~4 is due to evolution in the number of partially neutral cells, which affect the 21-cm in spite of their small volume fraction as the rest of the volume has zero brightness.  The flatness of the bispectra in boxes 4 and 5 clearly highlight the features introduced in the bispectrum by the neutral hydrogen regions required by the Ly$\alpha$ Forest.  In future 21-cm measurements, it would be valuable to correlate such features with the Ly$\alpha$ Forest data.

\section{Prospects of Detection} \label{sec:sensitivity}

For a preliminary study of the prospects of detecting the 21-cm bispectrum modelled above, we use the \21cms\footnote{\url{https://github.com/jpober/21cmSense}} \citep{2013AJ....145...65P,2014ApJ...782...66P} and \pyobs\ \citep{2022MNRAS.510.3838W} codes to model the bispectrum covariance induced by instrument noise for \skalow.  

Our assumed telescope parameters are the same as in \cite{RK21}. We have used the core configuration with 224 elements, with each element having diameter of 38~m and a field of view of about $12.5~\uni{deg}{2}$ at 150~MHz \citep{acedo2020ska}. The shortest and longest baselines for this configuration are 35.1~m and 887~m, respectively, resulting in an angular resolution of $10'$ at 150~MHz. 
We assume a tracking mode of 6~hr, at $z \sim 6$ ($\nu_o = 204$~MHz), for 180 days.  We include all $k$-modes in our calculation (using the `opt' foreground mode, i.e., assuming that the foregrounds are somehow removed from the data).  Using these noise estimates, we generate 100 noise boxes of $128^3$ resolution elements with different seeds and calculate their bispectra. In Figure~\ref{fig:sense1}, we show the variance of these noise bispectra and compare it with the signal at redshift $z=5.96$ for various $k$-modes\footnote{The low resolution ($128^3$) boxes do not have very large $k$-modes. Therefore, the last panel of Figure~\ref{fig:sense1} does not have bispectra at large $k_3$ values.}. We see that the bispectrum sensitivity is better at lower $k$-modes (larger scale). These results suggest that \skalow\ should be able to detect the 21-cm isosceles bispectra at small $k$-modes with $\sim1000~$hours of tracking mode observation, assuming optimistic foreground removal. These observations would help us understand the history of reionization, as well as the geometry of neutral islands at the end of reionization.

\section{Conclusions} \label{sec:conclusion}

We have computed the 21-cm bispectrum in simulations of cosmic reionization that are consistent with the observed \lya\ forest at $z > 5$.  Our findings can be summarised as follows:

\begin{itemize}
\item  A late end of reionization causes the 21-cm bispectrum and the neutral hydrogen fraction bispectrum to have large values at $z < 6$ (frequencies greater than 200~MHz).  This is in contrast to traditional reionization models, which typically predict zero bispectrum and power spectrum at these frequencies.
\item The neutral fraction bispectrum is negative at all scales during the early stages of reionization.  At later stages, the bispectrum starts to become positive at small scales (large $k$).  Towards the end of reionization, the bispectra are positive at all scales for squeezed and stretched limit triangles. But they remain negative around the equilateral configuration at small scales (large $k$). This suggests that most of the non-spherical, ``elongated'' features have the geometry of above-average regions (neutral islands) embedded in the below-average background (ionized regions), but there are still some small spherical below-average regions (ionization bubbles) embedded within above-average regions (neutral islands).
\item The 21-cm bispectrum follows the trends seen in the neutral fraction bispectrum.
\item For generic triangle configurations, the normalised bispectra of neutral fraction are negative at high redshifts. They then turn positive with the progress of reionization. The $\Delta T_\mathrm{b}$ bispectra show very similar behaviour, however, their amplitude is slightly more positive, reflecting the effect of the density bispectra. This effect of density is most readily observed in post-reionization brightness temperature bispectra, which are usually larger than the reionization bispectra at all $k$-mode triangles and do not show strong fluctuations with triangle index.
\item Partially ionised regions do not affect the shape of $x_\mathrm{HI}$ or 21-cm bispectra significantly. However, removing large ionised regions or neutral islands will completely change the shape, amplitude and sign of bispectra at any redshift.
\item With about a 1,000~hr of tracking mode observation and optimistic foreground removal, \skalow\ should be able to detect the bispectrum at $z\sim 6$ for small $k$-modes (large scale).
\end{itemize}

Overall, this work adds to the realization that statistics beyond the power spectrum can be very useful to understand reionization history from the future 21-cm observations. Similar to our previous work on the 21-cm power spectrum from the late reionization model, this work also highlights the importance of relatively higher frequency corresponding to redshifts $z<6$ for the late reionization. Combining these 21-cm bispectrum and power spectrum results with the {\it James Webb Space Telescope (JWST)} and {\it Nancy Grace Roman Space Telescope} should lead to further insights into reionization physics.

\section*{Acknowledgements}

We acknowledge useful discussions with Basudeb Dasgupta, Suman Majumdar, Rishi Khatri, Rajesh Mondal and Somnath Bharadwaj. GK gratefully acknowledges support by the Max Planck Society via a partner group grant.  This work used the Cambridge Service for Data Driven Discovery (CSD3) operated by the University of Cambridge (www.csd3.cam.ac.uk), provided by Dell EMC and Intel using Tier-2 funding from the Engineering and Physical Sciences Research Council (capital grant EP/P020259/1), and DiRAC funding from the Science and Technology Facilities Council (www.dirac.ac.uk).  This work further used the COSMA Data Centric system operated Durham University on behalf of the STFC DiRAC HPC Facility. This equipment was funded by a BIS National E-infrastructure capital grant ST/K00042X/1, DiRAC Operations grant ST/K003267/1 and Durham University. DiRAC is part of the UK's National E-Infrastructure. MH is supported by STFC (grant number ST/N000927/1). For the purpose of open access, the author has applied a Creative Commons Attribution (CC BY) licence to any Author Accepted Manuscript version arising from this submission. 

\section*{Data Availability}

Data shown in various figures will be made available upon reasonable request.

\bibliographystyle{mnras}

\appendix

\section{Resolution Tests} \label{sec:resolution}

\begin{figure*}
  \includegraphics[width=0.33\textwidth]{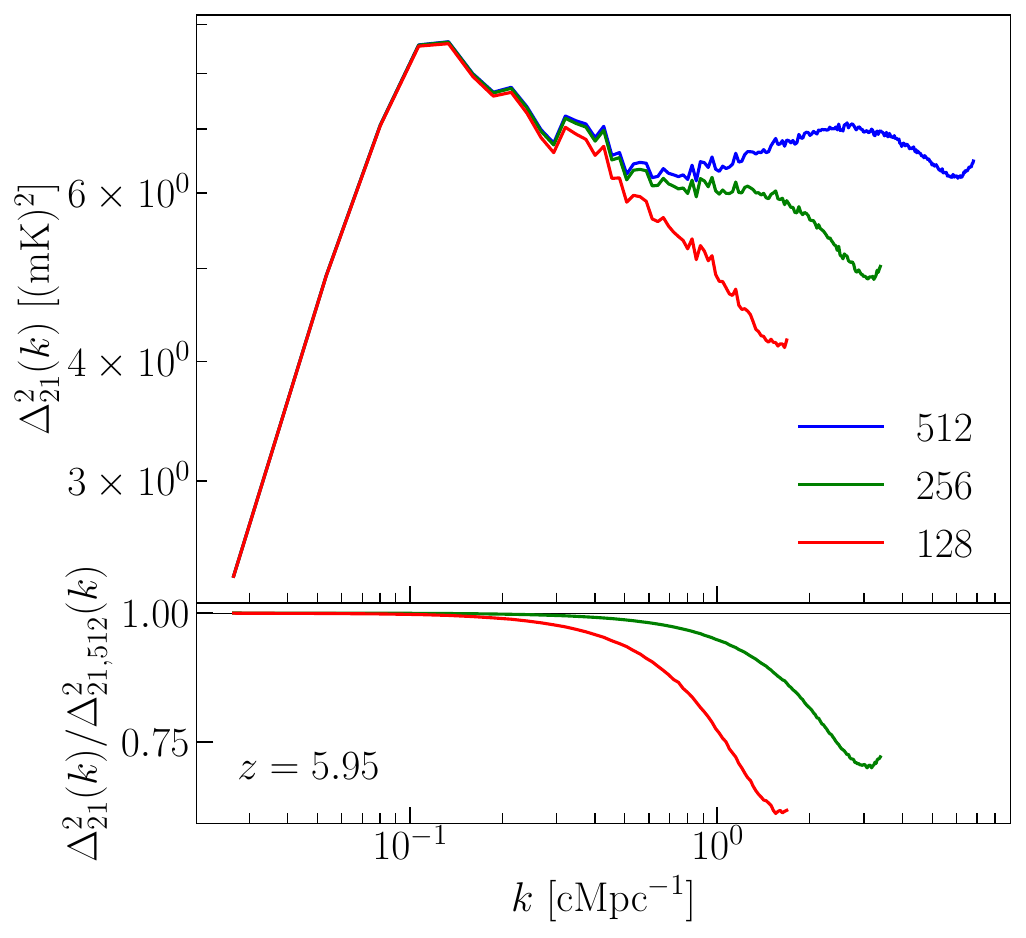}
  \includegraphics[width=0.327\textwidth]{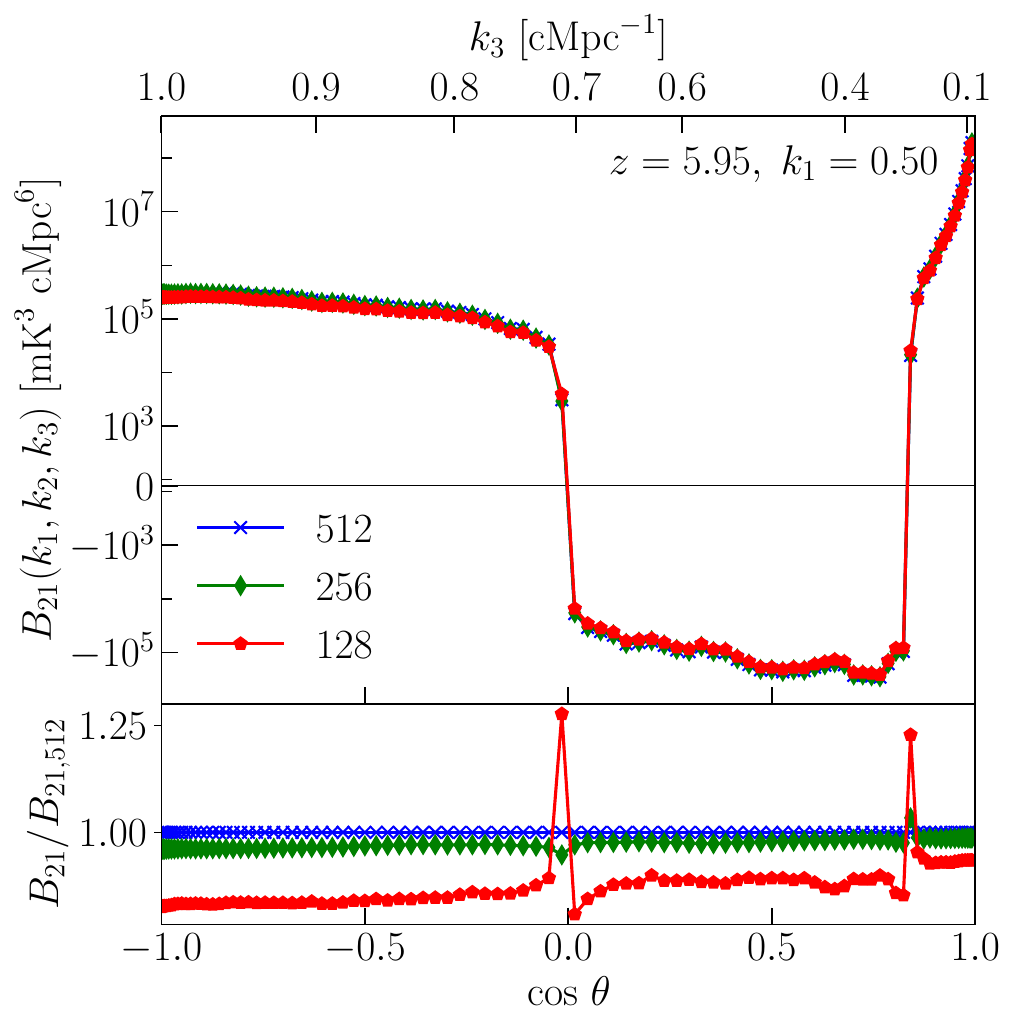}
  \includegraphics[width=0.333\textwidth]{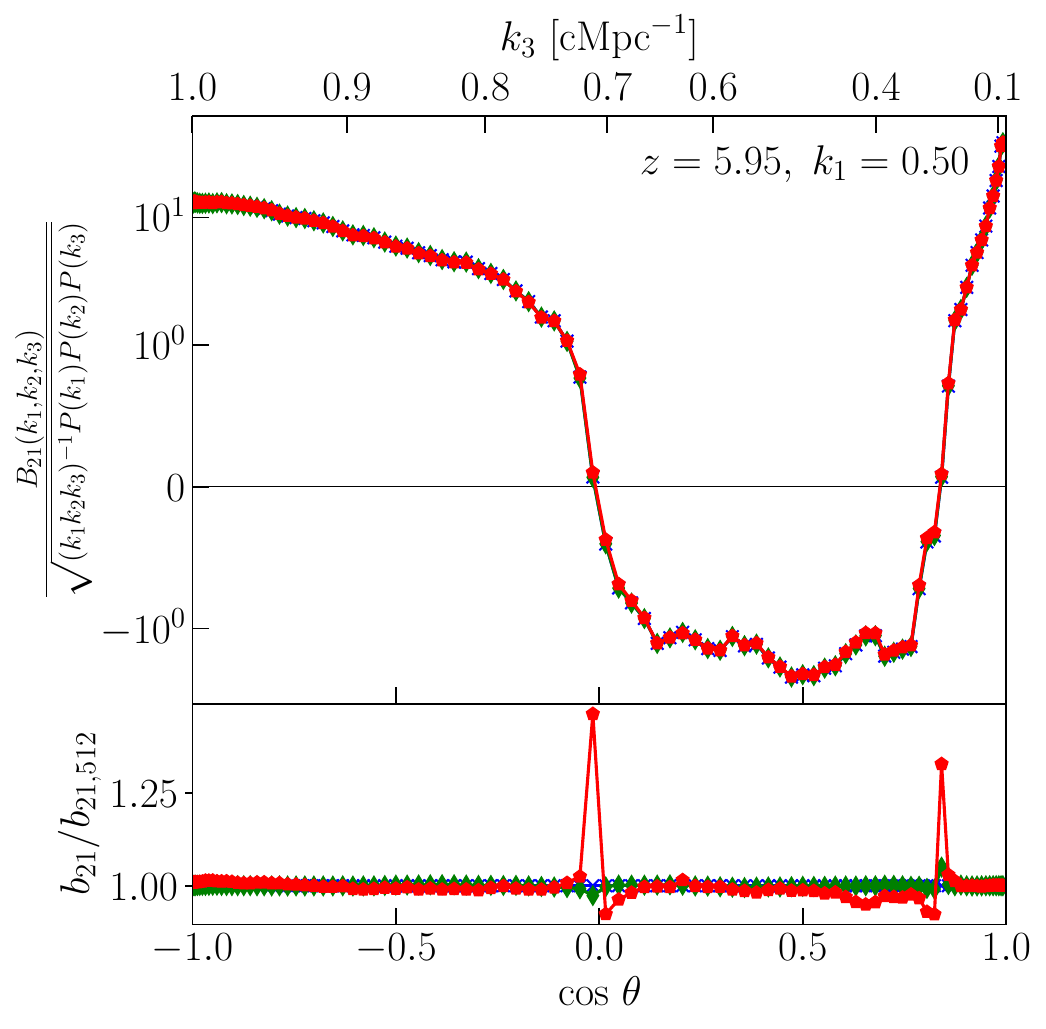}
  \caption{Effect of resolution on power spectrum (left), unnormalised (middle) and normalised (right) isosceles bispectrum with $k_1 = 0.5~{\rm cMpc}^{-1}$ at $z=5.96$. Both the power spectrum and unnormalised bispectrum shows suppression of power at large $k$-modes as we reduce the box resolution. However, their effects are cancelled in the normalised bispectrum, which appears robust with respect to box resolution. }
  \label{fig:res_1}
\end{figure*}

Our simulation boxes of the ionization, density and velocity have resolution of $2048^3$ pixels, therefore, our brightness temperature box also has the same resolution. However, it is difficult to calculate bispectrum for the full resolution boxes. Therefore, in our work, we have presented bispectrum for $256^3$ pixel box. We reduce the box resolution by averaging neighboring $n^3$ pixels into one large pixels, where $n>1$. In Figures~\ref{fig:res_1}, we show unnormalised and normalised isosceles bispectra at $k_1 = 0.5~{\rm cMpc}^{-1}$ and power spectra for $512^3$, $256^3$ and $128^3$ pixels resolutions at $z=5.96$ . 

As expected, the small scale (large $k$) power spectra diverge with resolution, as the fluctuations on these scales are averaged out while lowering the resolution. A similar trend can be seen for the unnormalised bispectra where the large $k$~modes show larger divergence for $128^3$ box. However, the normalised bispectra show very little fluctuations with box resolution as the effect of resolution on bispectrum and power spectrum cancel out due to normalisation. Note that the peaks in $128^3$ (red) box in bottom middle and right panels occur only at $k$-modes where the bispectra have very small amplitude. Overall, the results presented in our work are robust with box resolution at relevant scales.

\section{Traingle Indices} \label{sec:triangleindex}
In Table~\ref{tab:TriangleIndices}, we list the values of $(k_1, k_2, k_3)$ triplets for all triangle indices shown in Figure~\ref{fig:gen_1}. 

\begin{table*}
	\centering
	\caption{Values of $(k_1, k_2, k_3)$ triplets for all triangle indices shown in Figure~\ref{fig:gen_1}.  All $k$ values are in units of cMpc$^{-1}$.}
	\label{tab:TriangleIndices}
    \begin{tabular}{c|c|c|c|c|c|c|c|c|c|c|c}
	\toprule
	Triangle Index & $k_1$ & $k_2$ & $k_3$ 	& Triangle Index & $k_1$ & $k_2$ & $k_3$ & Triangle Index & $k_1$ & $k_2$ & $k_3$ \\
	\hline
		1  &  0.100  &  0.100  &  0.075  &  54  &  1.100  &  1.000  &  0.100  &  107  &  1.000  &  0.750  &  0.372  \\ 
		2  &  0.200  &  0.200  &  0.075  &  55  &  0.750  &  0.750  &  0.129  &  108  &  0.750  &  0.500  &  0.457  \\ 
		3  &  0.500  &  0.500  &  0.090  &  56  &  0.200  &  0.200  &  0.131  &  109  &  0.750  &  0.750  &  0.467  \\ 
		4  &  0.108  &  0.100  &  0.100  &  57  &  0.500  &  0.500  &  0.163  &  110  &  0.503  &  0.500  &  0.500  \\ 
		5  &  0.126  &  0.100  &  0.100  &  58  &  1.000  &  1.000  &  0.163  &  111  &  0.590  &  0.500  &  0.500  \\ 
		6  &  0.150  &  0.100  &  0.100  &  59  &  0.204  &  0.200  &  0.200  &  112  &  0.710  &  0.500  &  0.500  \\ 
		7  &  0.168  &  0.100  &  0.100  &  60  &  0.240  &  0.200  &  0.200  &  113  &  0.811  &  0.500  &  0.500  \\ 
		8  &  0.183  &  0.100  &  0.100  &  61  &  0.287  &  0.200  &  0.200  &  114  &  0.893  &  0.500  &  0.500  \\ 
		9  &  0.199  &  0.100  &  0.100  &  62  &  0.327  &  0.200  &  0.200  &  115  &  0.974  &  0.500  &  0.500  \\ 
		10  &  0.204  &  0.100  &  0.100  &  63  &  0.359  &  0.200  &  0.200  &  116  &  0.997  &  0.500  &  0.500  \\ 
		11  &  0.200  &  0.110  &  0.100  &  64  &  0.391  &  0.200  &  0.200  &  117  &  1.000  &  0.501  &  0.500  \\ 
		12  &  0.200  &  0.110  &  0.100  &  65  &  0.401  &  0.200  &  0.200  &  118  &  1.000  &  0.515  &  0.500  \\ 
		13  &  0.200  &  0.118  &  0.100  &  66  &  0.500  &  0.303  &  0.200  &  119  &  1.000  &  0.549  &  0.500  \\ 
		14  &  0.200  &  0.142  &  0.100  &  67  &  0.500  &  0.306  &  0.200  &  120  &  0.750  &  0.664  &  0.500  \\ 
		15  &  0.200  &  0.178  &  0.100  &  68  &  0.500  &  0.317  &  0.200  &  121  &  1.000  &  0.666  &  0.500  \\ 
		16  &  0.200  &  0.199  &  0.100  &  69  &  0.500  &  0.361  &  0.200  &  122  &  0.764  &  0.750  &  0.500  \\ 
		17  &  0.229  &  0.200  &  0.100  &  70  &  0.500  &  0.438  &  0.200  &  123  &  0.903  &  0.750  &  0.500  \\ 
		18  &  0.253  &  0.200  &  0.100  &  71  &  0.500  &  0.480  &  0.200  &  124  &  1.024  &  0.750  &  0.500  \\ 
		19  &  0.274  &  0.200  &  0.100  &  72  &  0.541  &  0.500  &  0.200  &  125  &  1.120  &  0.750  &  0.500  \\ 
		20  &  0.296  &  0.200  &  0.100  &  73  &  0.596  &  0.500  &  0.200  &  126  &  1.217  &  0.750  &  0.500  \\ 
		21  &  0.301  &  0.200  &  0.100  &  74  &  0.640  &  0.500  &  0.200  &  127  &  1.247  &  0.750  &  0.500  \\ 
		22  &  0.500  &  0.403  &  0.100  &  75  &  0.686  &  0.500  &  0.200  &  128  &  1.000  &  0.868  &  0.500  \\ 
		23  &  0.500  &  0.403  &  0.100  &  76  &  0.699  &  0.500  &  0.200  &  129  &  1.000  &  0.972  &  0.500  \\ 
		24  &  0.500  &  0.409  &  0.100  &  77  &  0.750  &  0.552  &  0.200  &  130  &  1.120  &  1.000  &  0.500  \\ 
		25  &  0.500  &  0.425  &  0.100  &  78  &  0.750  &  0.555  &  0.200  &  131  &  1.250  &  1.000  &  0.500  \\ 
		26  &  0.500  &  0.461  &  0.100  &  79  &  0.750  &  0.565  &  0.200  &  132  &  1.357  &  1.000  &  0.500  \\ 
		27  &  0.500  &  0.481  &  0.100  &  80  &  0.750  &  0.601  &  0.200  &  133  &  1.463  &  1.000  &  0.500  \\ 
		28  &  0.512  &  0.500  &  0.100  &  81  &  0.750  &  0.674  &  0.200  &  134  &  1.496  &  1.000  &  0.500  \\ 
		29  &  0.542  &  0.500  &  0.100  &  82  &  0.750  &  0.716  &  0.200  &  135  &  1.000  &  0.750  &  0.593  \\ 
		30  &  0.566  &  0.500  &  0.100  &  83  &  0.778  &  0.750  &  0.200  &  136  &  1.000  &  1.000  &  0.621  \\ 
		31  &  0.592  &  0.500  &  0.100  &  84  &  0.835  &  0.750  &  0.200  &  137  &  0.752  &  0.750  &  0.750  \\ 
		32  &  0.600  &  0.500  &  0.100  &  85  &  0.883  &  0.750  &  0.200  &  138  &  0.884  &  0.750  &  0.750  \\ 
		33  &  0.750  &  0.651  &  0.100  &  86  &  0.934  &  0.750  &  0.200  &  139  &  1.062  &  0.750  &  0.750  \\ 
		34  &  0.750  &  0.653  &  0.100  &  87  &  0.949  &  0.750  &  0.200  &  140  &  1.215  &  0.750  &  0.750  \\ 
		35  &  0.750  &  0.657  &  0.100  &  88  &  1.000  &  0.801  &  0.200  &  141  &  1.338  &  0.750  &  0.750  \\ 
		36  &  0.750  &  0.673  &  0.100  &  89  &  1.000  &  0.804  &  0.200  &  142  &  1.459  &  0.750  &  0.750  \\ 
		37  &  0.750  &  0.706  &  0.100  &  90  &  1.000  &  0.814  &  0.200  &  143  &  1.495  &  0.750  &  0.750  \\ 
		38  &  0.750  &  0.727  &  0.100  &  91  &  1.000  &  0.848  &  0.200  &  144  &  1.000  &  0.903  &  0.750  \\ 
		39  &  0.758  &  0.750  &  0.100  &  92  &  1.000  &  0.918  &  0.200  &  145  &  1.050  &  1.000  &  0.750  \\ 
		40  &  0.788  &  0.750  &  0.100  &  93  &  1.000  &  0.959  &  0.200  &  146  &  1.252  &  1.000  &  0.750  \\ 
		41  &  0.814  &  0.750  &  0.100  &  94  &  1.021  &  1.000  &  0.200  &  147  &  1.425  &  1.000  &  0.750  \\ 
		42  &  0.842  &  0.750  &  0.100  &  95  &  1.080  &  1.000  &  0.200  &  148  &  1.564  &  1.000  &  0.750  \\ 
		43  &  0.850  &  0.750  &  0.100  &  96  &  1.130  &  1.000  &  0.200  &  149  &  1.703  &  1.000  &  0.750  \\ 
		44  &  1.000  &  0.901  &  0.100  &  97  &  1.182  &  1.000  &  0.200  &  150  &  1.744  &  1.000  &  0.750  \\ 
		45  &  1.000  &  0.903  &  0.100  &  98  &  1.198  &  1.000  &  0.200  &  151  &  1.002  &  1.000  &  1.000  \\ 
		46  &  1.000  &  0.906  &  0.100  &  99  &  0.750  &  0.750  &  0.239  &  152  &  1.178  &  1.000  &  1.000  \\ 
		47  &  1.000  &  0.922  &  0.100  &  100  &  0.750  &  0.500  &  0.256  &  153  &  1.416  &  1.000  &  1.000  \\ 
		48  &  1.000  &  0.955  &  0.100  &  101  &  1.000  &  0.750  &  0.256  &  154  &  1.619  &  1.000  &  1.000  \\ 
		49  &  1.000  &  0.975  &  0.100  &  102  &  0.750  &  0.500  &  0.271  &  155  &  1.783  &  1.000  &  1.000  \\ 
		50  &  1.006  &  1.000  &  0.100  &  103  &  1.000  &  0.750  &  0.288  &  156  &  1.945  &  1.000  &  1.000  \\ 
		51  &  1.036  &  1.000  &  0.100  &  104  &  0.500  &  0.500  &  0.312  &  157  &  1.993  &  1.000  &  1.000  \\ 
		52  &  1.063  &  1.000  &  0.100  &  105  &  0.750  &  0.500  &  0.317  &    &    &    &    \\ 
		53  &  1.091  &  1.000  &  0.100  &  106  &  1.000  &  1.000  &  0.317  &    &    &    &    \\ 
	\bottomrule
	\end{tabular}
\end{table*}

\bsp	
\label{lastpage}
\end{document}